\documentclass[prb,aps,twocolumn,amsmath,amssymb,floatfix,
superscriptaddress]{revtex4-2}
\usepackage[dvips]{graphics}
\usepackage[justification=raggedright]{caption}
\usepackage{subcaption}
\usepackage[font=footnotesize,figurewithin=none]{caption}
\usepackage{subcaption}
\usepackage{adjustbox}
\usepackage[labelfont=bf]{caption}
\usepackage[labelformat=parens, labelfont=bf]{subcaption}
\usepackage{color}
\usepackage{float}
\usepackage{natbib}
\usepackage[dvipsnames]{xcolor}
\setcitestyle{numbers}
\setcitestyle{square}
\definecolor{dred}{rgb}{0,0,0.6}
\usepackage{graphicx}    
\usepackage{bm}  
\usepackage{amssymb} 
\usepackage{mathtools}
\usepackage{mathdots}
\usepackage{amsmath}
\usepackage{braket}
\usepackage[mathscr]{euscript}
\usepackage[colorlinks=true, allcolors=blue]{hyperref}
\usepackage{url}
\begin{document}
\title{Single and multi-frequency driving protocols in a Rashba nanowire proximitized to an \textit{s}-wave superconductor}% Force line breaks with \\

\author{Koustav Roy and Saurabh Basu \\ \textit{Department of Physics, Indian Institute of Technology Guwahati-Guwahati, 781039 Assam, India}}
%\email{saurabh@iitg.ac.in}
%\author{Tapan Mishra, Saurabh Basu}%
% %\email{}
%\affiliation{%

\date{\today}% It is always \today, today,
\begin{abstract}
We perform systematic analyses of single and multi-frequency driving protocols on a Rashba nanowire with superconducting correlations induced by proximity effects. The results for the single-mode drive reveal interesting frequency dependencies of the Majorana modes, in the sense that the parameters corresponding to the trivial and the topological limits of the undriven (static) case host Majorana zero modes, respectively at low and high frequencies. Further, emergence of long-range interactions are noted that give rise to multiple gap-closing scenarios, where the latter implies occurrence of multiple Majorana modes. On the other hand, the multi-frequency driving protocol, sub-grouped into commensurate and incommensurate ratios of the frequencies, demonstrates intriguing consequences. The commensurate case yields dynamical control over the stability of the edge modes. Whereas, more intricate driving protocols, such as those with larger frequency ratios are studies as well and they are found to harm the Majoranas by pushing them into the bulk. Finally, the incommensurate case yields independent Majorana modes occurring at low-symmetry points in the Brillouin zone. While the single and the commensurate multi-frequency driving protocols admit the usage of symmetric time frames for the computation of the topological invariants, the incommensurate case relies on the framework of many-mode Floquet theory, where the topological properties are ascertained via calculating the Berry phase. We present band structure and phase diagrams to substantiate all our results. The robustness and concurrent existence of these unique Majorana modes, even amidst a very dense energy spectrum along with a lack of global time-periodicity, hold promise for future applications in the field of quantum computation.
\end{abstract}

\maketitle

%\tableofcontents
\begin{center}\section{\label{sec:level1}Introduction}\end{center}

Ever since the discovery of integer quantum Hall effect \cite{iqhe}, the study of topological states of matter has signaled a pivotal advancement in solid-state research \cite{topology1,topology2,topology3,topology4,topology5}. Exploring and manipulating states that are topologically protected has received substantial interest both theoretically \cite{topologytheory1,topologytheory2,topologytheory3} and experimentally \cite{topologyexpt1,topologyexpt2,topologyexpt3,topologyexpt4}. In particular and of relevance to us, topological superconductors (TSCs) have garnered significant attention due to the emergence of exotic quasiparticle excitations akin to Majorana fermions \cite{Majoranafermion}. These Majorana bound states (MBS) associated with TSCs offer a compelling avenue for non-abelian braiding, and conveniently crucial for achieving fault-tolerant quantum computation \cite{qubit1,qubit2,qubit3}. The simplest prototype for achieving topological superconductivity was initially proposed by Kitaev, utilizing a one-dimensional model of spinless $p$-wave superconductors \cite{kitaevchain}. However, the scarcity of $p$-wave superconductors in nature has hindered the experimental realization of Kitaev's model for TSCs. Nonetheless, the seminal contributions of Fu and Kane \cite{fukane1,fukane2} have shown that a Kitaev chain can be realized in various systems, utilizing the conventional proximity-induced $s$-wave superconductivity as a key ingredient. Building on their insights, the experimental accessibility of the Kitaev chain has expanded significantly, leading to the proposal of MBS in diverse media such as one-dimensional semiconducting nanowire with Rashba spin orbit coupling (SOC) \cite{kitaevexpt1,kitaevexpt2,kitaevexpt3,kitaevexpt4}, two-dimensional quantum wells \cite{kitaevexpt5,kitaevexpt6}, carbon nanotubes \cite{kitaevexpt7,kitaevexpt8} and so on.
\par Further, the investigation of Majorana states can be extended beyond equilibrium conditions. Driven systems are particularly valuable in this context, as they reveal a broader spectrum of topological phenomena than their static counterpart. A key aspect of our research is the examination of Majoranas in driven systems which may be beneficial for advancing prospects of quantum information processing. Notably, the driven Kitaev chain has been proposed as a promising platform for demonstrating topologically protected, non-Abelian Majorana braiding operations within a single nanowire \cite{Floquet_braiding1,Floquet_braiding2}.  This has led to significant interest in Floquet engineering, a technique that allows precise tuning of topological properties in non-equilibrium conditions \cite{floquetformalism1,floquetformalism2,floquetformalism3}. Such systems are particularly intriguing because they can host localized edge modes at energy $\pi/T$ (also termed as Majorana $\pi$ modes) in addition to the conventional Majorana zero modes \cite{floquet1,floquet2,floquet3,floquet4}.
The growing interest in this field is further fueled by experimental progress that validates these novel characteristics. For instance, localized Floquet Majoranas have been observed in experiments with cold atom quantum wires \cite{FMOBS1,quantum_wire_MBS} and 2D cold atom superfluids \cite{FMOBS2}, where the chemical potential is periodically modulated. Furthermore, it has been shown that driven nanowires can support unpaired Majoranas even without a magnetic field \cite{FM_without_magnetic_field}. Also, recent experiments on planar Josephson junctions have reported differential conductance ($dI/dV$) measurements that suggest the presence of multiple Andreev reflections that are indicative of these exotic states \cite{josephson1,josephson2,josephson3,josephson4,josephson5}. 
% Moreover, the concept of generating non-trivial Floquet characteristics has been effectively utilized in numerous experiments involving ultracold atoms in optical lattices out-of-equilibrium conditions and photonic waveguides \cite{ultracoldfloquet1,ultracoldfloquet2,ultracoldfloquet3,ultracoldfloquet4,ultracoldfloquet5,ultracoldfloquet6}. 
In recent years, there has been a remarkable surge in interest for studying non-equilibrium topology of systems in the following sense. Is topology robust in a driven scenario? In fact, surprisingly, topology not only survives but gets richer in presence of periodic drive. This encompasses various phenomena, such as the generation of higher winding or Chern numbers in 1D, 2D, and quasi-1D systems \cite{1dfloquet1,1dfloquet2,1dfloquet3,creutzfloquet,chern1,chern2,paolo2}, emergence of discrete time crystalline phases along with period doubling oscillations \cite{period2t1,period2t2}, topological characterization of quantum chaos models \cite{dkr,khm}, emergence of Floquet-Anderson phases in quasiperiodic systems \cite{dimerizedkitaevfloquet} and so forth. 
\par In the pursuit of understanding non-equilibrium topology, researchers have explored Floquet topological superconductors to reveal dynamical versions of Majoranas \cite{paolo2,kitaevfloquet1,kitaev_high_freq,kitaevfloquet2,kitaevfloquet3,kitaevfloquet4,kitaevfloquet5,odd_freq_cayao}. However, much of this research has been based on Kitaev's original model involving $p$-wave superconductors. Nevertheless, considering the practicality of experimentation, our attention will pivot towards the Rashba nanowire model, as the exploration of Floquet Majoranas within this model remains relatively uncharted territory,  with only a few studies conducted so far \cite{rashbafloquet1,rashbafloquet2}. In contrast to the original model, the Rashba nanowire variant incorporates $s$-wave superconducting correlations and a magnetic field that breaks the time-reversal symmetry (TRS). The lack of TRS constrains the number of Majorana modes since the formation of many Majorana modes demands the protection of TRS \cite{kitaevfloquet1,TRSkitaev}. This motivates us to look for a scenario whether an external periodic drive (entering through the magnetic field itself) in the present context could admit the emergence of multiple Majorana modes which may have crucial ramifications in topological quantum computation.
\par Furthermore, in the context of external periodic driving, few fundamental question arises. What are the implications for the generalization of topological properties when the external drive includes more than one frequency? Can we anticipate the coexistence of multiple Majorana modes? If so, then how can we adequately characterize them? In a generic sense, there can be two possible alternatives for multi-frequency driving: (i) when the ratio of the two frequencies is a rational number (commensurate drive), and (ii) when it is an irrational number (incommensurate drive). Recent advancements in Floquet engineering have spurred considerable progress in this area. Specifically, the formalism associated with incommensurate driving expands the dimensionality of the problem by introducing a Fourier manifold corresponding to each frequency \cite{floquetincommensurate1,floquetincommensurate2,floquetincommensurate3}. This treatment of creating synthetic dimensions has been extensively explored, yielding practical applications, such as, zero-dimensional qubit mixers \cite{zerodqubit}, energy converters between different drives \cite{floquetincommensurate1,floquetincommensurate4}, and stabilizers for dynamic topological phases \cite{floquetincommensurate5}. However, due to the necessity of additional Fourier dimensions, this approach may pose computational challenges for models exceeding two dimensions. On the contrary, commensurate driving methods, notably two-tone harmonic driving, have found significant application in a range of experimentally achievable settings. These include instances such as quantum destructive interference in the Fermi-Hubbard model \cite{fermihubbard}, sensitivity synchronizer in Rydberg atoms \cite{rydberg}, generation of Floquet-Bloch non-trivial band structures \cite{solin,paolomulti}, etc. Drawing inspiration from these advancements, in this paper along with the single-drive scenario, we seek to examine the topological characteristics of the Rashba nanowire model under a multi-frequency (both commensurate and incommensurate) driving protocols.
\begin{figure}[t]
    \begin{subfigure}[b]{\columnwidth}
         \includegraphics[width=\columnwidth]{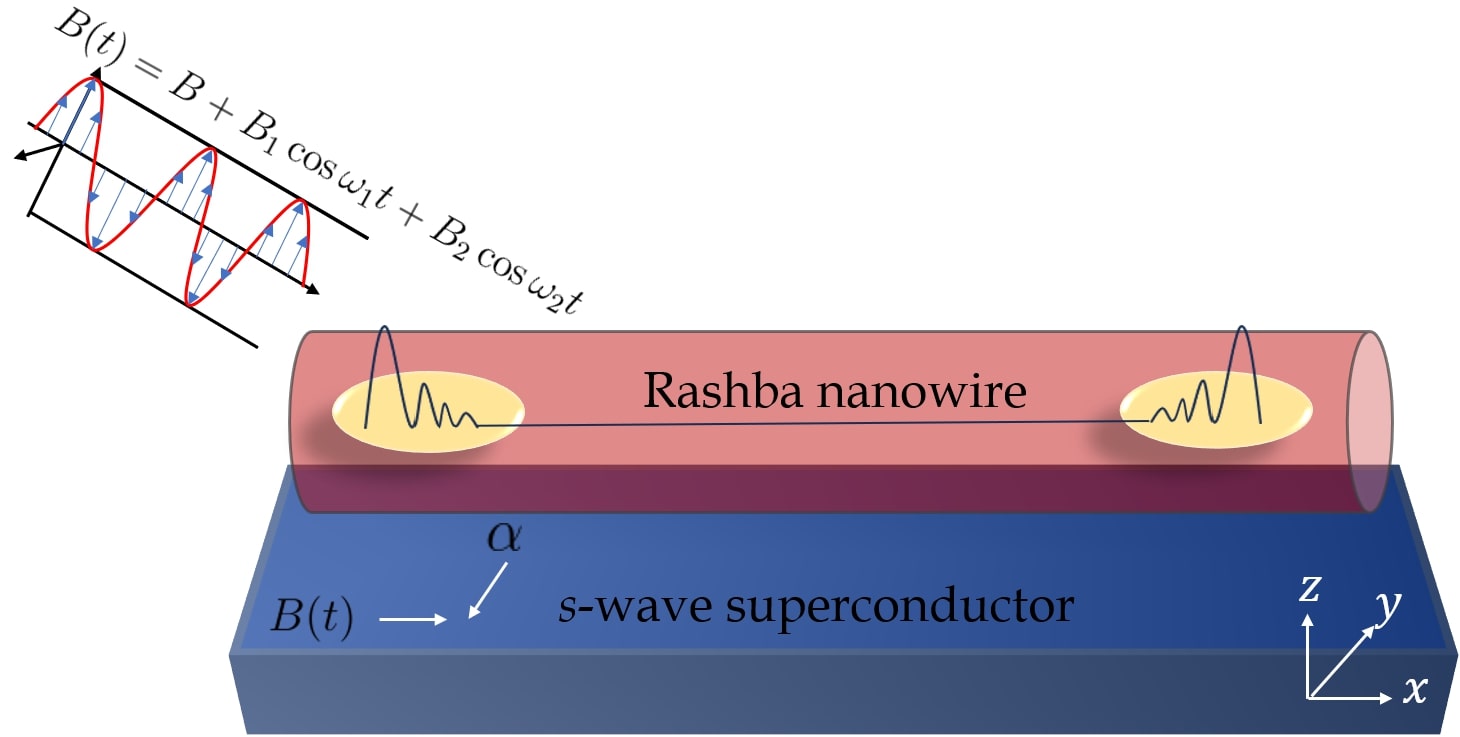}
        %  \caption{}
         \label{1}
     \end{subfigure}
\caption{{The figure depicts a schematic representation of a Rashba nanowire proximity-coupled to an $s$-wave superconductor subjected to a perpendicular time-modulated in-plane magnetic field, $B(t)$ applied in the $x$-direction.}} 
\label{1}
\end{figure}
\par Our findings are derived from the mapping of the drives into the frequency space, resulting in the emergence of additional Floquet couplings. Consequently, a thorough analysis of frequencies may lead to various instances of non-trivial gap closure, ensuring sustained topological nature of the system even for parameter values that host trivial phases. Furthermore, we aim to utilize the symmetries associated with the stroboscopic time evolution operator, thereby introducing a new approach to the topological classification, where each non-trivial phase is distinguished by the concept of \emph{`pairs of symmetric time frames'}. Furthermore, within a commensurate driving protocol, there exists a possibility to dynamically adjust the stability and the number of Majorana edge modes by carefully tuning the ratio of driving frequencies. Conversely, in presence of an incommensurate driving scenario, the absence of global time periodicity impedes our traditional technique of using an evolution operator and instead demands the usage of many-mode Floquet formalism illustrated in Ref.~\cite{multimode1} and others \cite{multimode2,multimode3,validityofmultimode}. This methodology may lead to the emergence of unique Majorana modes that remain stable and can coexist independently with other Majoranas (from the static picture), despite the dense spectra induced by the quasiperiodic nature of the drive.
\par The layout of the subsequent discussions is as follows: In Sec.~\ref{sec:level2} we provide an overview of the static version of the model, revisiting its symmetries and the functionalities of each of the parameters. Following this, we shall introduce the Floquet tool to formulate an effective time-independent Hamiltonian. In Sec. \ref{sec:level3}, we explore the topological properties associated with a single-drive scenario and its topological classification. In Sec. \ref{sec:level4} we cover multifrequency commensurate and incommensurate drives, discussing the relevant formalism and the results. We also propose a potential experimental setup to validate our findings. Finally, in Sec.~\ref{sec:level5}, we provide a summary and draw conclusions based on our findings.
\vspace{0.3in}
\begin{center}{\section{\label{sec:level2}The \textit{s}-wave Kitaev chain and the Floquet Formalism}}\end{center}
A potential physical realization for engineering topological superconductivity in one dimension based on Kitaev's model is schematically depicted in Fig.~\ref{1}. This setup involves the utilization of a semiconductor nanowire, such as InSb, proximitized to an $s$-wave superconductor where a strong Rashba SOC, $\alpha$ plays a crucial role and is an important ingredient of our study. The corresponding Hamiltonian including an in-plane (along $x$-direction) magnetic field, $B$ can be written as,
\begin{equation}
    H = \hat{\mathcal{H}}_{kin} + \hat{\mathcal{H}}_{so} + \hat{\mathcal{H}}_z + \hat{\mathcal{H}}_{sc},
\label{Ham_static}
\end{equation}
where,
\begin{align}
    \hat{\mathcal{H}}_{so} &= \alpha \sum_j \left[\hat{c}^\dagger_{j, \downarrow} (\hat{c}_{j+1,\uparrow} - \hat{c}_{j-1,\uparrow}) - \hat{c}^\dagger_{j,\uparrow} (\hat{c}_{j+1,\downarrow} - \hat{c}_{j-1,\downarrow})\right], \label{Eq:SOC} \\
    \hat{\mathcal{H}}_z &= \sum_j B \left(\hat{c}^\dagger_{j,\uparrow} \hat{c}_{j,\downarrow} + \hat{c}^\dagger_{j,\downarrow} \hat{c}_{j,\uparrow}\right), \label{Eq:Zeeman} \\
    \hat{\mathcal{H}}_{sc} &= \frac{1}{2} \sum_j \left[\Delta \left(\hat{c}^\dagger_{j,\uparrow} \hat{c}^\dagger_{j,\downarrow} - \hat{c}^\dagger_{j,\downarrow} \hat{c}^\dagger_{j,\uparrow}\right) + h.c.\right], \label{Eq:SC}
\end{align}
with $\Delta$ being the amplitude of $s$-wave pairing and, the usual kinetic energy for the particles in the nanowire given by,
% \begin{equation}
% %\begin{split}
%     \hat{\mathcal{H}}_{so} = \alpha \sum_j \Big[\hat{c} ^\dagger _{j, \downarrow} ( \hat{c}_{j+1,\uparrow} - \hat{c}_{j-1,\uparrow}) - \hat{c}^\dagger _{j,\uparrow}( \hat{c}_{j+1,\downarrow} - \hat{c}_{j-1,\downarrow})\Big],
%     \end{equation}
% \begin{equation}
%    \hat{\mathcal{H}}_z = \sum_j B ( \hat{c}^\dagger _{j,\uparrow} \hat{c}_{j,\downarrow} + \hat{c}^\dagger _{j, \downarrow} \hat{c} _{j,\uparrow} ) ,
%    \label{Ham_zeeman}
% \end{equation}
% \begin{equation}
%     \hat{\mathcal{H}}_{sc} = \frac{1}{2} \sum_j \Big[ \Delta ( \hat{c}^\dagger _{j,\uparrow} c^\dagger _{j,\downarrow}- \hat{c}^\dagger _{j,\downarrow} c^\dagger _{j,\uparrow}) + h.c \Big],
% \end{equation}
\begin{equation}
    \hat{\mathcal{H}}_{kin} = -\mu \sum_{j,\sigma} ^N \hat{c}^\dagger _{j,\sigma} \hat{c}_{j,\sigma} + \sum_{j,\sigma}^ {N-1} \Big [-t_{hop} \hat{c}^\dagger _{j,\sigma} \hat{c} _{j+1,\sigma} + h.c.\Big ].
\end{equation}
To provide a visual representation, let us decompose each element of the Hamiltonian to explore their physical significance.
\begin{figure}[t]
    \begin{subfigure}[t]{\columnwidth}
         \includegraphics[width=\columnwidth]{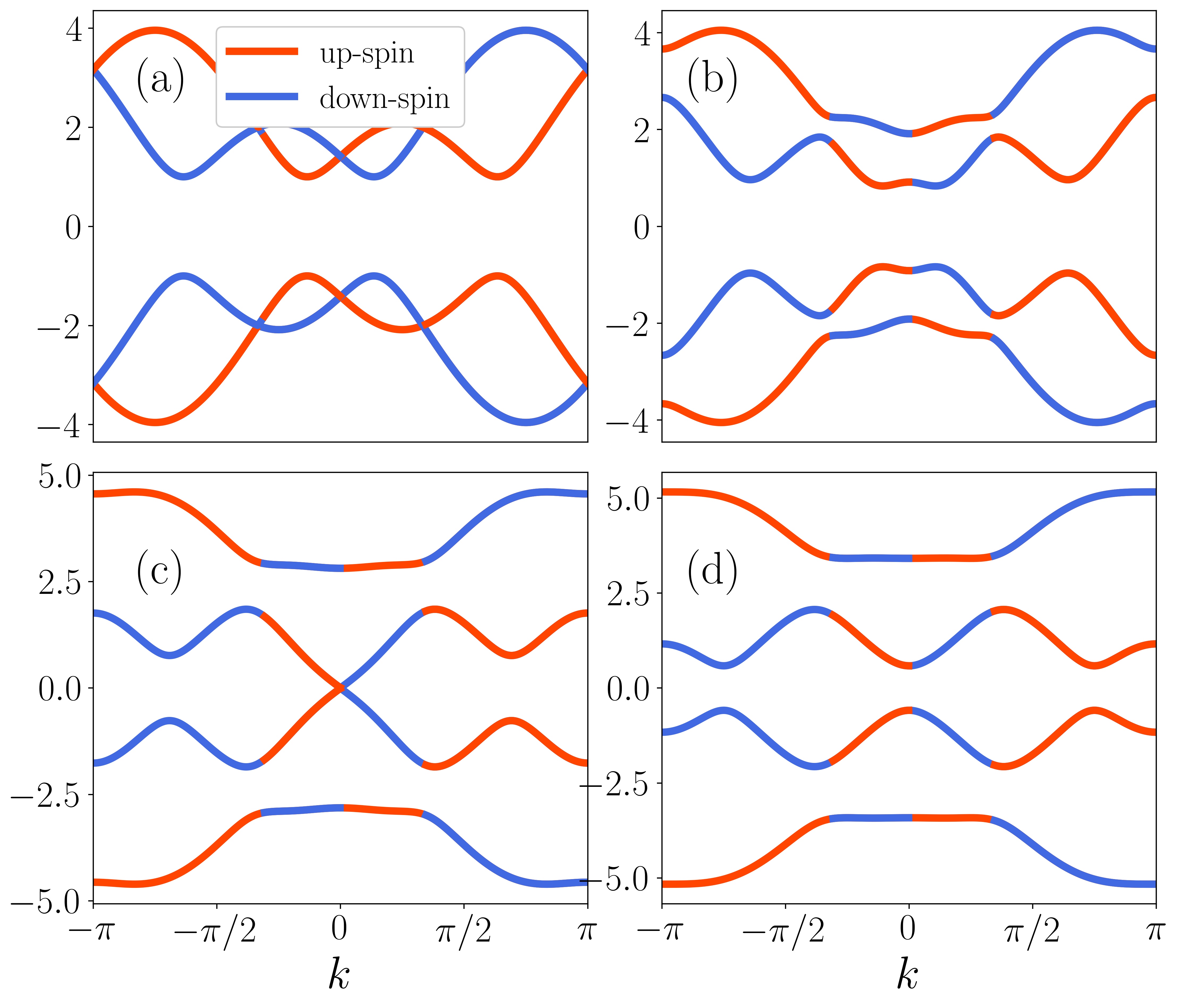}
        %  \caption{}
         \label{Fig2}
     \end{subfigure}
\caption{{The dispersion spectra corresponding to a static $s$-wave Kitaev chain for different values of the magnetic field, $B$. We study the case of zero field (a), 0.5 $B_{c_1}$ (b), $B_{c_1}$ (c), and $1.5B_{c_1}$ (d). A gap-closing transition (at $k=0$) occurring at $B=B_{c_1}$ is noted. The other parameter are chosen as, $\mu=0.5,\Delta=0.5,t_{hop}=1,$ and $\alpha=0.5$.}} 
\label{2}
\end{figure}  
The inherent particle-hole symmetry of the model necessitates the condition $\gamma^{\dagger} = \gamma$  (where $\gamma$ is the Majorana operator). This implies that the existence of a particle being its own antiparticle must occur at zero energy. However, in the case of $s$-wave superconductors, the presence of spin degrees of freedom prevents the existence of robust zero-energy modes. To address this limitation, one may introduce an in-plane magnetic field $B$ shown in Eq.~\ref{Eq:Zeeman}, such that, if the Hamiltonian for particles is denoted by $H(B)$, then for holes it would be $-H(-B)$.
Additionally, considering the presence of spin, there has to be a Zeeman coupling that will change sign under TRS. Another crucial aspect that must be taken into account is that $s$-wave superconductors exhibit singlet pairing, implying that the total spin of every excitation is conserved. Consequently each singlet excitation possesses a definite spin, including the Majoranas. However, this poses a threat to the existence of Majoranas, which should have no definite spin due to their property of being complex conjugate of themselves. Therefore, the only viable solution is to violate spin conservation by introducing a Rashba SOC ($\alpha$) (shown in Eq.~\ref{Eq:SOC}), which effectively breaks the spin conservation.
% \begin{figure}[t]
%           \begin{subfigure}[!h]{0.493\columnwidth}
%          %\centering
%          \includegraphics[width=\columnwidth]{Picture19.png}
%          \captionlistentry{}
%          \label{2.1}
%      \end{subfigure}
%      \begin{subfigure}[!h]{0.493\columnwidth}
%          \includegraphics[width=\columnwidth]{Picture20.png}
%          \captionlistentry{}
%          \label{2.2}
%      \end{subfigure}
%      \begin{subfigure}[!h]{0.493\columnwidth}
%          \includegraphics[width=\columnwidth]{Picture21.png}
%          \captionlistentry{}
%          \label{2.3}
%      \end{subfigure}
%      \begin{subfigure}[!h]{0.493\columnwidth}
%          \includegraphics[width=\columnwidth]{Picture22.png}
%          \captionlistentry{}
%          \label{2.4}
%      \end{subfigure}
% \caption{{The dispersion spectra corresponding to a static $s$-wave Kitaev chain for different values of the magnetic field, $B$. We study the case of zero field (a), 0.5 $B_c$ (b), $B_c$ (c), and $1.5B_c$ (d). A gap-closing transition (at $k=0$) occurring at $B=B_c$ is noted. The other parameter are chosen as, $\mu=0.5,\Delta=0.5,t=1,$ and $\alpha=0.5$.}} 
%  \label{Fig2}
% \end{figure}
\begin{figure*}[t] % Use figure* to span both columns
\begin{minipage}{\linewidth}
        \begin{subfigure}[t]{0.33\columnwidth}
            \includegraphics[width=\linewidth]{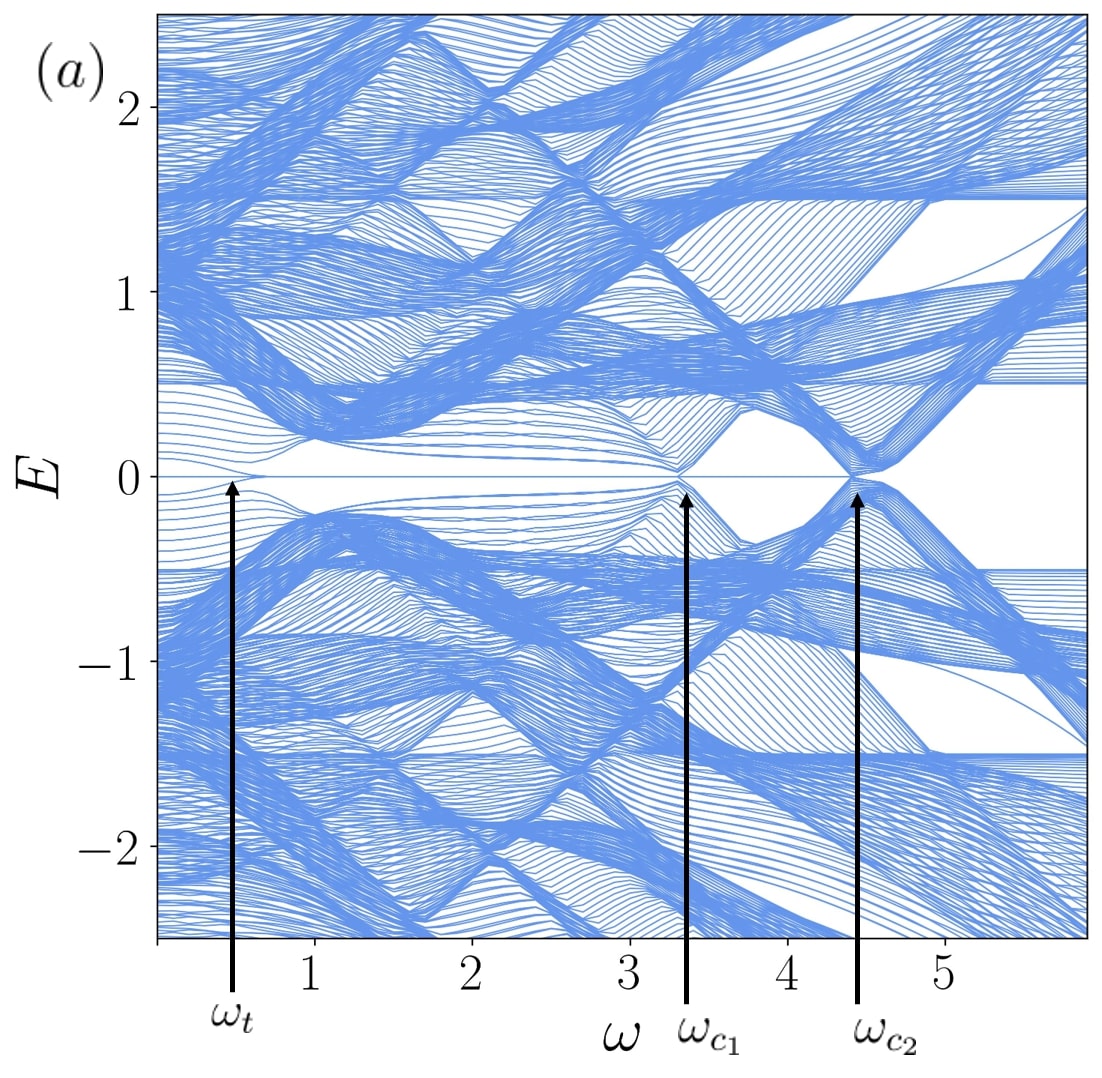}
            %\caption{}
        \end{subfigure}%
        \hfill
        \begin{subfigure}[t]{0.33\columnwidth}
            \includegraphics[width=\linewidth]{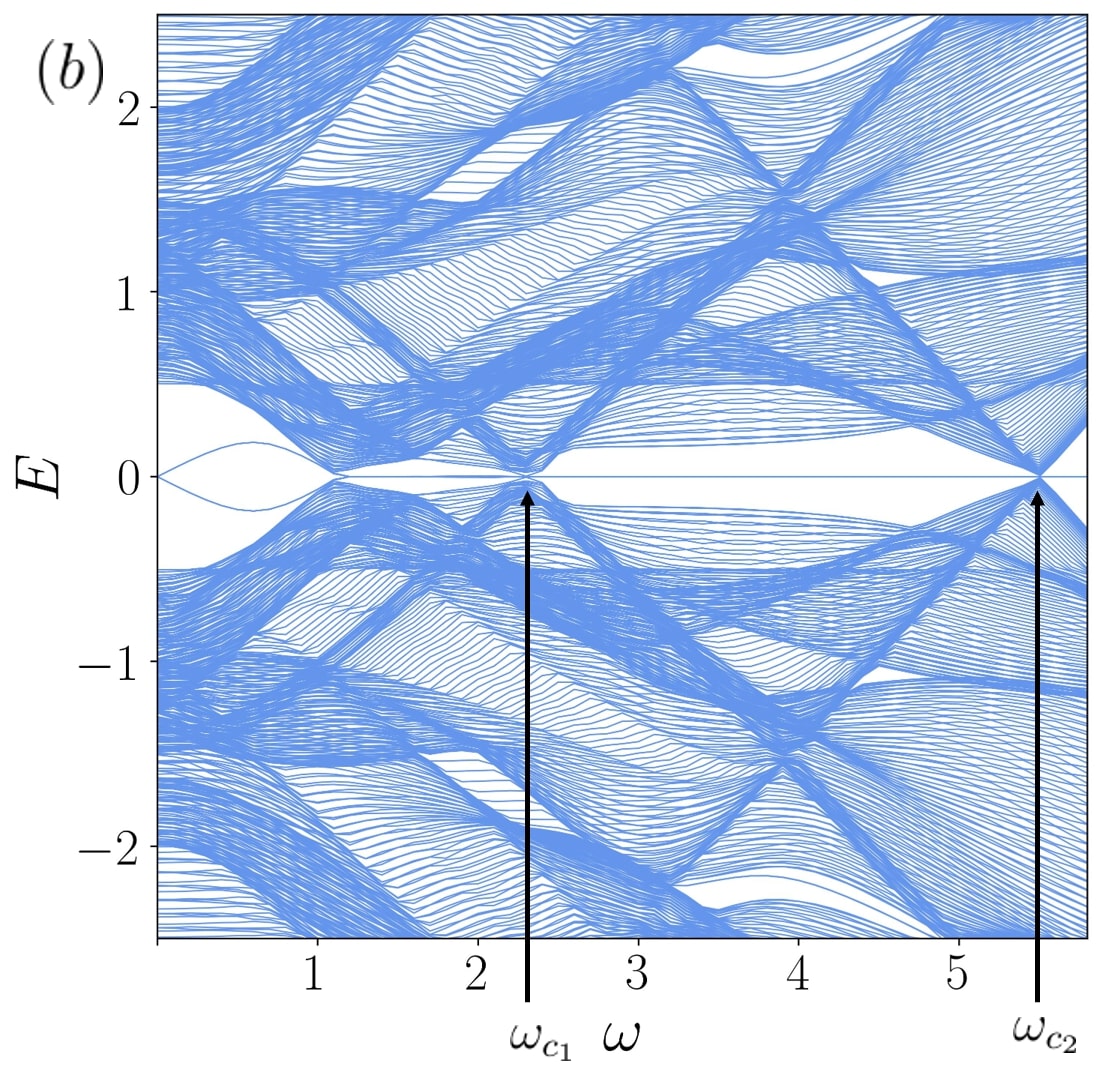}
            %\caption{}
        \end{subfigure}%
        \hfill
        \begin{subfigure}[t]{0.33\columnwidth}
            \includegraphics[width=\linewidth]{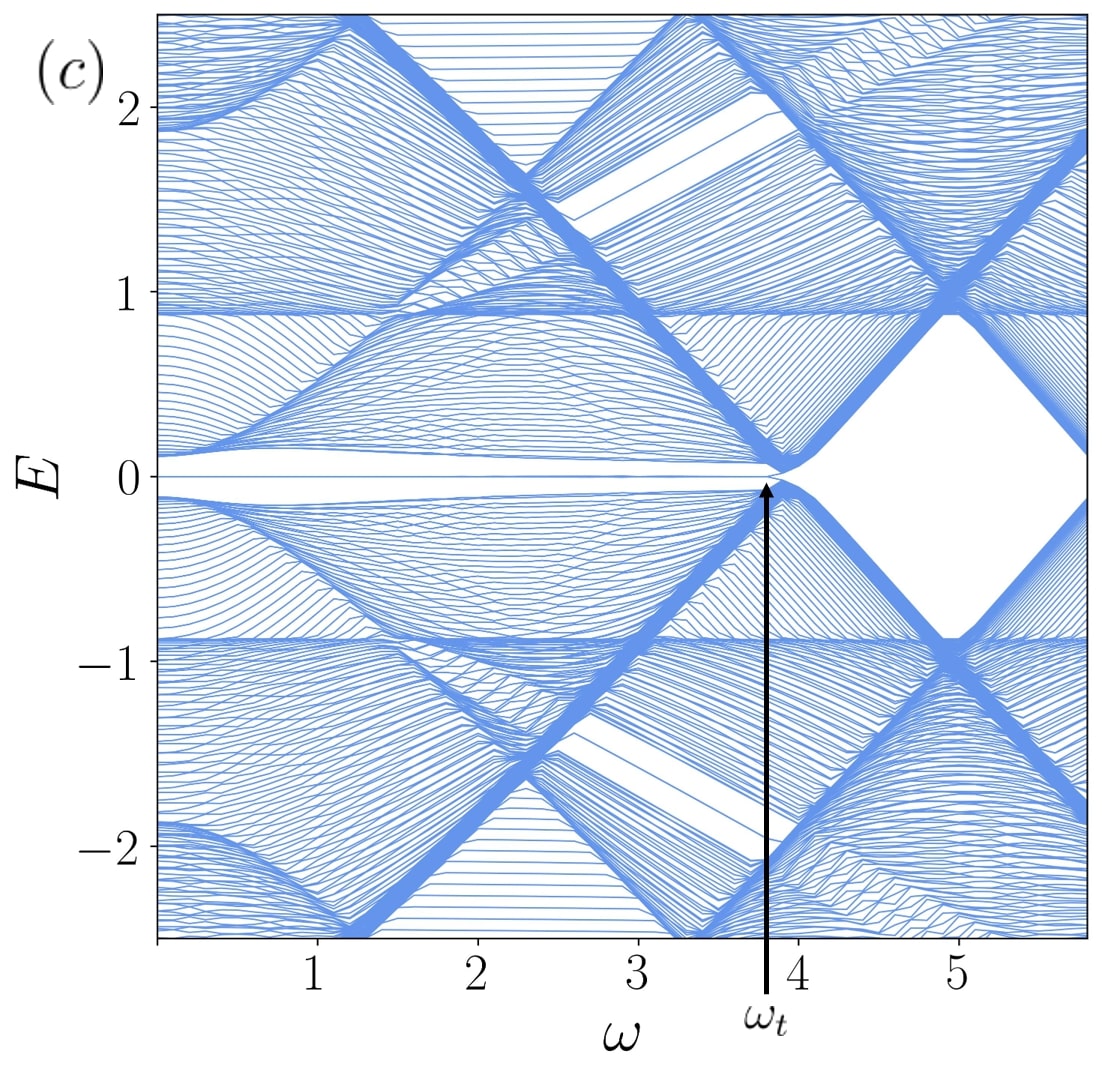}
            %\caption{}
        \end{subfigure}
\caption{The figure depicts real space quasi-energy spectrum (with $m \in [-1,1]$) in the Shirley-Floquet representation plotted as a function of frequency for different magnetic fields. Panel (a) and (c) represent the static trivial limits, $B<B_{c_1}$ and $B>B_{c_2}$ respectively. Whereas, panel (b) represents the static topological condition, $B_{c_1}<B<B_{c_2}$. The value of the magnetic fields are chosen as, $B=0.5$ in panel (a), $B=1.5$ in panel (b) and $B=5$ in panel (c). The rest of the parameters are chosen as, $t_{hop}=1,\mu=0,\Delta=1,B_1=1$.}
\label{Fig3}
    \end{minipage}
\end{figure*}
\par Finally, in the tight binding notations of the Bogoliubov de Gennes (BdG) representation in the Fourier space, the Hamiltonian assumes the form,
\begin{equation}
\begin{split}
&H=\sum_{k}\psi^\dagger(x)\mathcal{H}(k)  \psi(x)dx\\
& \mathcal{H}(k)=[(2t_{hop}-\mu)-2t_{hop} \cos{k}] \tau_z \sigma_0 + 2 \alpha \sin{k} \tau_z \sigma_y  \\ &  \qquad ~ ~ ~ ~ + B \tau_z \sigma_x + \Delta \tau_y \sigma_y\\
& \psi ^\dagger = (c^\dagger _{k,\uparrow} , c^\dagger _{k,\downarrow} , c_{k,\uparrow} , c_{k,\downarrow}).
\end{split}
\end{equation}
Here, $\psi_{k,\uparrow} (\psi_{k,\uparrow}^{\dagger})$ and $\psi_{k,\downarrow} (\psi_{k,\downarrow}^{\dagger})$ represent electron annihilation (creation) operators for the spin up and spin down sectors, and $\sigma,\tau$ denote the spin and particle degrees of freedom, respectively. 
\par At this point, it is crucial to address the symmetries inherent to the model. As previously discussed, the Hamiltonian upholds particle-hole symmetry, with $C = \tau_x \sigma_0 \kappa$ ($\kappa$ is the complex conjugation operator) as the particle-hole operator one finds, $C^{-1} \mathcal{H}(k) C = -\mathcal{H}(-k)$. However, it violates time-reversal symmetry, where $T = i \tau_0 \sigma_y \kappa$ yields $T^{-1} \mathcal{H}(k) T \neq \mathcal{H}(-k)$. Consequently, the model is classified under symmetry class D, and characterized by a $\mathbb{Z}_2$ invariant that should host at most one pair of Majorana modes \cite{10fold}.
% \begin{widetext}
%     \begin{minipage}{\linewidth}
%         \begin{figure}[H]
%         %\centering
%         \hspace{-8mm}
%           \begin{subfigure}[!h]{0.33\columnwidth}
%          %\centering
%          \includegraphics[width=\columnwidth]{Figure_3a.jpg}
%          \caption{$B<B_{c_1}$}
%          \label{Fig3.1}
%      \end{subfigure}
%      \begin{subfigure}[!h]{0.33\columnwidth}
%          \includegraphics[width=\columnwidth]{Figure_3b.jpg}
%          \caption{$B_{c_1}<B<B_{c_2}$}
%          \label{Fig3.2}
%      \end{subfigure}
%      \begin{subfigure}[!h]{0.33\columnwidth}
%          \includegraphics[width=\columnwidth]{Figure_3c.jpg}
%          \caption{$B>B_{c_2}$}
%          \label{Fig3.3}
%      \end{subfigure}
% \caption{{The figure depicts real space quasi-energy spectrum (with $m \in [-1,1]$) in the Shirley-Floquet representation plotted as a function of frequency for different magnetic fields. Panel (a) and (c) represent the static trivial limits, $B<B_{c_1}$ and $B>B_{c_2}$ respectively. Whereas, panel (b) represents the static topological condition, $B_{c_1}<B<B_{c_2}$. The value of the magnetic fields are chosen as, $B=0.5$ in panel (a), $B=1.5$ in panel (b) and $B=5$ in panel (c). The rest of the parameters are chosen as, $t=1,\mu=0,\Delta=1,B_1=1$.}} 
%  \label{Fig3}
% \end{figure}    
% \end{minipage}
% \end{widetext}
\par We review the results of the static model in order to set the stage for discussing results corresponding to the driven situation. At $k=0$, the Hamiltonian undergoes a gap-closing transition for a particular value of the magnetic field, namely, $B=B_{c_1}=\sqrt{\mu^2 + \Delta^2}$, Further another transition occurs at $k=\pm \pi$ for $B=B_{c_2}=\sqrt{(\mu-4t_{hop})^2 + \Delta^2}$. These phenomena of bulk gap closing aid in defining the boundaries of the topological phases,
as illustrated in Fig.~\ref{2}, which showcases the spectral properties of the bulk. In Fig.~\ref{2}a, the dispersion at zero field $(B = 0)$ is depicted. It is evident that there are two distinct bands corresponding to different spins: one for particles (at negative energies) and the other for holes (at positive energies). The pairing term, $\Delta$ induces a gap between these bands, while the SOC term, $\alpha$ shifts the two spin bands horizontally. As soon as a magnetic field is switched on with $B < B_{c_1}$, for example, $B = 0.5B_{c_1}$ (see Fig.~\ref{2}b), the bands do not cross, thereby opening gaps in the spectrum. With further increment in the value of $B$, at $B = B_{c_1}$, the outermost bands move further away, whereas the gap between the inner bands closes (see Fig.~\ref{2}c). As the Zeeman field exceeds this critical value, $B_{c_1}$ the spectrum becomes fully gapped again (Fig.~\ref{2}d). This phenomenon of gap opening and closing signifies a topological phase transition occurring at $B = B_{c_1}$, with the topological phase persisting in the region $B_{c_1}<B<B_{c_2}$. Hnece, the system is topologically non-trivial in the range, $B_{c_1}<B<B_{c_2}$, whereas, topologically trivial in the range, $B<B_{c_1}$ and $B>B_{c_2}$ respectively. All these results are consistent with previously reported findings \cite{review_on_Majorana1,review_on_Majorana2,review_on_Majorana3} and thus lay the groundwork for exploring the driven systems. 
\par With the above groundwork being rolled out, we can now delve into the Floquet topological characteristics of our periodically driven model, where the time-dependent modulation of the magnetic field plays a central role. Specifically, we have examined a harmonic drive applied to the magnetic field. Unlike in static scenarios, where the system remains in the eigenstate
at each instantaneous time, the driven system may absorb quantized energy from the external fields, featuring non-equilibrium
properties. As a result, the Floquet version of the system can reveal more intricate topological features compared to its static counterpart. Furthermore, a prerequisite in Floquet engineering is the theoretical approach to compute the effective Hamiltonian, derived from the stroboscopic evolution of the system. However, given the harmonic nature of the drive, it is advantageous to carry out our analysis in the frequency domain, which is analogous to the `dressed atom' picture of the electromagnetic field-induced atomic systems \cite{dressed_atom}. This approach allows us to get access to the localization of Majorana modes in both topologically non-trivial and trivial (static) regimes. To initiate our analysis, we consider a sinusoidal drive represented by the Hamiltonian $H(t) = H_0 + V_{0}\cos{\omega t}$, where $V_0$ and $\omega$ denote the amplitude and frequency of the driving term, respectively, and $H_0$ is a generic static Hamiltonian. Floquet theory then allows us to solve the time-dependent Schrödinger equation using the Floquet ansatz $\ket{\psi(t)} = e^{-i E t} \ket{u(t)}$, where $\ket{u(t+T)}=\ket{u(t)}$ denotes the time-periodic Floquet modes, and $E$ represents the Floquet quasi-energies. Similar to quasi-momentum in crystals, these quasi-energies can be thought of as a periodic variable defined in the Floquet Brillouin zone (FBZ), $E \in [-\pi/T:\pi/T]$. On the other hand, the Floquet modes can also be interpreted as the eigenstates of the Floquet stroboscopic time evolution operator via,
\begin{equation}
\begin{split}
    & \hat{U}(T) \ket{\psi(0)} = \ket{\psi(T)}, \\ & \hat{U}(T) \ket{u(0)} = e^{-iET} \ket{u(T)} = e^{-iET} \ket {u(0)},
\end{split}
\end{equation}
with the Floquet stroboscopic time evolution operator being,
\begin{equation}
    \hat{U}(T) = \mathcal{T} \text{exp} [-i \int_{0}^{T} H(t) dt] =  e^{-iH_{\text{eff}}T}.
    \label{evolutionoperator}
\end{equation}
Here $\mathcal{T}$ denotes the time ordering product and $H_{\text{eff}}$ is the effective time-independent Hamiltonian. One can obtain $E$ and $\ket{u(t)}$ by solving the Floquet-Bloch equation,
\begin{equation}
[H(t)-i\partial_t]\ket{u_k(t)}= E \ket{u_k(t)}.
\end{equation}
The operator $H(t)-i\partial_t = H_F$ is termed as Floquet Hamiltonian. Because of the time periodicity, it is convenient
to consider the composite Hilbert space $\mathscr{R}\otimes \mathscr{T}$ where $\mathscr{R}$ is the usual Hilbert space with a complete set of orthogonal basis, and $\mathscr{T}$ is the space of time periodic functions spanned by $e^{-im\omega t}$. This yields the following form of $H_F$,
\begin{equation}
H_F=\sum_{m,m^\prime} \Big( m\omega \delta_{m,m^{\prime}} + \frac{1}{T} \int_{0}^{T} dt H(t) e^{-i(m-m^\prime)\omega t} \Big).
\label{Eq:SF1}
\end{equation} 
This leads to a situation where we can split the driven spectrum into an infinite number of copies of the undriven Hamiltonian separated by $m\omega$ where, the index $m$ defines a subspace, called as the $m^{th}$ Floquet replica. A general representation of the Floquet Hamiltonian thus can be represented as,
\begin{equation}
%\hspace*{-0.7cm}
\label{Eq:SF2}
    H_{F} = \scriptsize{\begin{bmatrix}\ddots & \vdots  & \vdots & \vdots &\vdots & \vdots & \iddots \\ \dots & H_0 - 2\omega & H_{-1} & H_{-2} & H_{-3} & H_{-4} & \dots \\ \dots & H_1 & H_0  - \omega & H_{-1} & H_{-2} & H_{-3} & \dots \\ \dots & H_2 & H_1 & H_0 & H_{-1} & H_{-2} & \dots \\ \dots & H_3 & H_2 & H_1 & H_0 +\omega & H_{-1} & \dots \\
    \dots & H_4 & H_3 & H_2 & H_1 & H_0 + 2\omega & \dots \\ \iddots & \vdots & \vdots & \vdots & \vdots & \vdots & \ddots
    \end{bmatrix}},
\end{equation}
where the elements $H_{\pm m}=\frac{1}{T}\int^T_0 {H(t)e^{\pm i m \omega t} dt}$ get rid of the explicit time dependence. The Hamiltonian $H_F$ is termed as the Shirley-Floquet (SF) Hamiltonian. 
Now considering a harmonic drive, associated with the magnetic field, say, $H(t) = H_0 + \mathcal{H}_z(t)$, where, 
\begin{equation}
    \hat{\mathcal{H}}_z (t) =  \Big( B + B_1 \cos{\omega_1 t} \Big) \sum_j \Big [\hat{c}^\dagger _{j,\uparrow} \hat{c}_{j,\downarrow} + \hat{c}^\dagger _{j, \downarrow} \hat{c} _{j,\uparrow} \Big ],
\end{equation} and $H_0$ being the static part of the Hamiltonian, the only surviving terms in Eq.~\ref{Eq:SF1} and Eq.~\ref{Eq:SF2} are those with $m=\pm1$. Consequently, Eq.~\ref{Eq:SF1} can be rewritten as,
\begin{equation}
    \bra{m} H_F \ket{m^{\prime}} = m \omega \delta_{m,m^{\prime}} + E_0 \delta_{m,m^{\prime}} + \mathcal{H}^{m-m^{\prime}},
    \label{Eq:SF3}
\end{equation}
where, $E_0$ is the energy of the static Hamiltonian and the off-diagonal terms in Eq.~\ref{Eq:SF3} in the $m,m^{\prime}$ basis are expressed as,
\begin{equation}
    \mathcal{H}^{m-m^{\prime}} = B_{1} /2 \sum_j \Big [\hat{c}^\dagger _{j,\uparrow} \hat{c}_{j,\downarrow} + \hat{c}^\dagger _{j, \downarrow} \hat{c} _{j,\uparrow} \Big ] \Big( \delta_{m-m^{\prime},1} + \delta_{m-m^{\prime},-1}\Big).
\end{equation}
Note that, an important property of these Floquet eigenstates is that they are localized in $m$, decaying rapidly for $|m-m_0|\omega \gg B_1$, where $m_0$ is the center of a given localized state. This
localization in frequency space is analogous to the well-known localization of Wannier-Stark states in real space \cite{WS_localization1,WS_localization2}, characterized by a localization length, $[\ln(\omega/V)]^{-1}$ with $V < \omega$, where $V$ denotes the nearest neighbor hopping amplitude. Consequently, the SF formalism can be thought of as the motion of a particle hopping in a one-dimensional synthetic lattice, spanned by the coordinate $m$, with $\omega$ acting as a uniform force field also known as Wannier-Stark field.
\par In the next section, we shall incorporate this method to further study the driven version of our model.
\begin{figure}[t]
    \begin{subfigure}[b]{\columnwidth}
         \includegraphics[width=0.75\columnwidth]{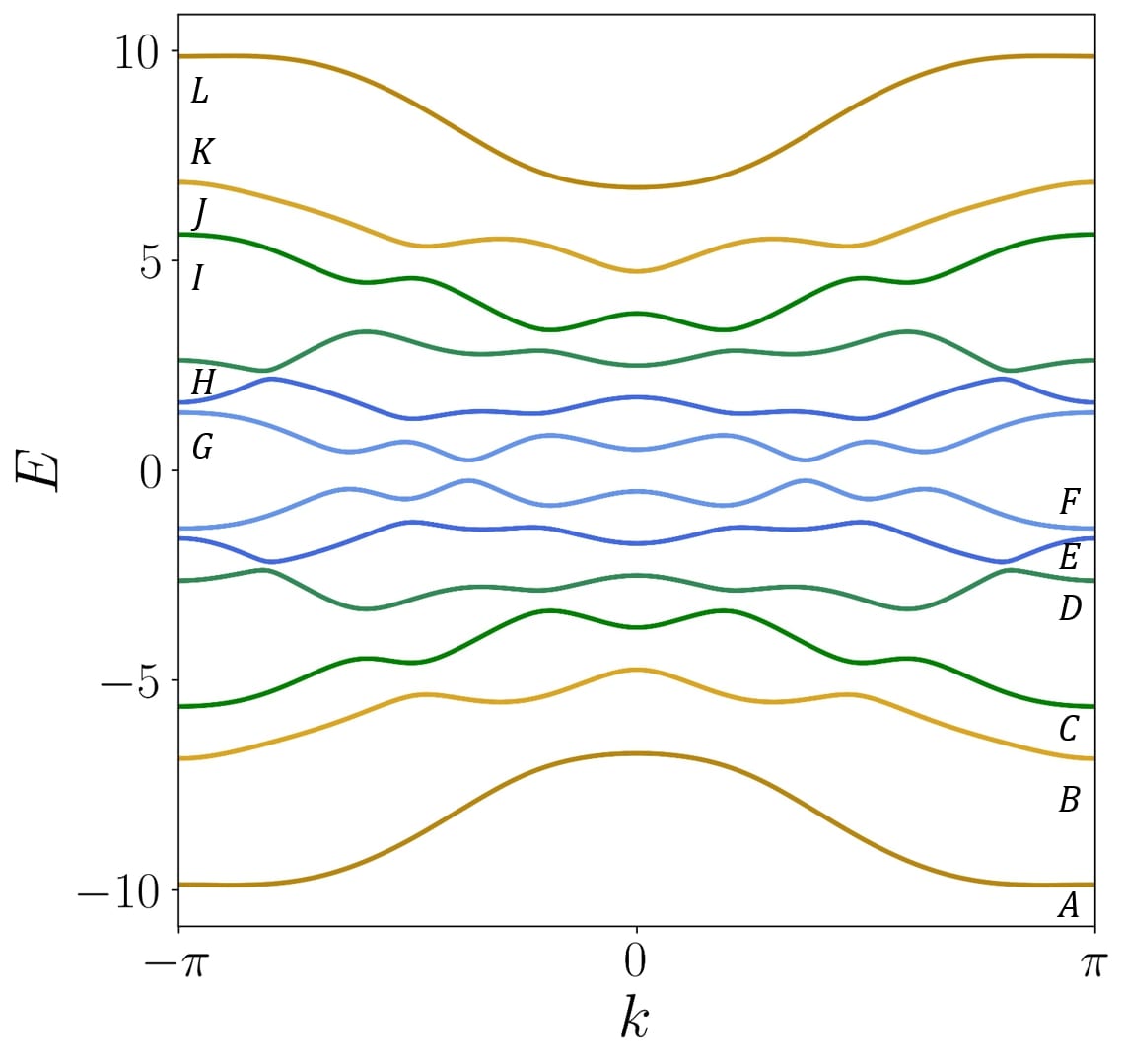}
        %  \caption{}
         \label{Fig4}
     \end{subfigure}
\caption{{The figure depicts bulk quasi-energy spectrum (with $m \in [-1,1]$) in the Shirley-Floquet representation. The different bands are denoted with letters $A-L$. The rest of the parameters are chosen as, $B_1=1$, $\omega=4$, and $B=1.5$ (static topological limit).}} 
\label{Fig4}
\end{figure}
\begin{center}{\section{\label{sec:level3}Single-frequency drive}}\end{center}
We first describe a harmonic drive, associated with the magnetic field which takes the form of a bi-chromatic waveform such that Eq.~\ref{Eq:Zeeman} can be rewritten as,
\begin{equation}
   \hat{\mathcal{H}}_z (t) =  \Big( B + B_1 \cos{\omega_1 t} + B_2 \cos{\omega_2 t} \Big) \sum_j \Big [\hat{c}^\dagger _{j,\uparrow} \hat{c}_{j,\downarrow} + \hat{c}^\dagger _{j, \downarrow} \hat{c} _{j,\uparrow} \Big ]. 
   \label{Ham_zeeman_periodic} 
\end{equation}
The rest of the terms in Eq.~\ref{Ham_static} are left unaltered. The single drive scenario can be exploited with the conditions, $B_2 = 0$ and $\omega_1 = \omega$. The Fourier components $|H|_{\pm m}$, except for $m=0,\pm1$ vanish owing to the mathematical form of the drive. Furthermore, by definition, the Floquet Hamiltonian is exact if one considers an infinite number of replicas. However, for numeric computation, we have to consider a finite number of them. The number of such replicas under consideration is decided by the
strength of the driving frequency, $\omega$. Suppose, the driving frequency is of the order or greater than the bandwidth, that is, $\omega \ge D$, ($D$ being the bandwidth), different admixtures of the replicas with $m > \pm 1$ can be neglected. Hence, we can truncate the infinite dimensional matrix into a block and can study the corresponding quasi-energy spectrum. By using Floquet theory, we can show that driving induces additional gaps and edge states depending upon the driving frequency and the strength of the driving field. To provide evidence of a topological phase transition in our driven scenario, it is essential to examine the real-space Floquet quasi-energy spectrum across a wide range of frequencies. For benchmarking, Fig.~\ref{Fig3} illustrates the driven quasi-energy spectrum in both trivial and topological situations, corresponding to the static case. Throughout this paper, we have standardized the energy unit as $t_{hop}$, the hopping amplitude. The other parameters are chosen as, $\alpha = 1, \Delta = 1, \mu = 0, B_1 = 1$. Upon switching on the time-dependent perturbation, both in the $B<B_{c_1}$ and $B>B_{c_2}$ regions where the system was entirely trivial in the static case, non-trivial behavior emerges in the presence of driving, accompanied by the appearance of Majorana zero modes (MZMs). Additionally, Majorana $\pi$ modes (MPMs), absent in the static case also manifest. At present, our focus is on the zero-energy modes depicted in Fig.~\ref{Fig3} to highlight the distinctions between static trivial and topological limits. Furthermore, multiple gap-closing transitions can be observed, indicating the potential generation of many Majorana modes, which we shall subsequently validate using the topological invariants. By analytically solving the Hamiltonian, we can get three critical values for the driving frequency corresponding to which gap-closing transition occurs. They are,
\begin{figure}[!t]
\centerline{\hfill
\includegraphics[width=0.25\textwidth]{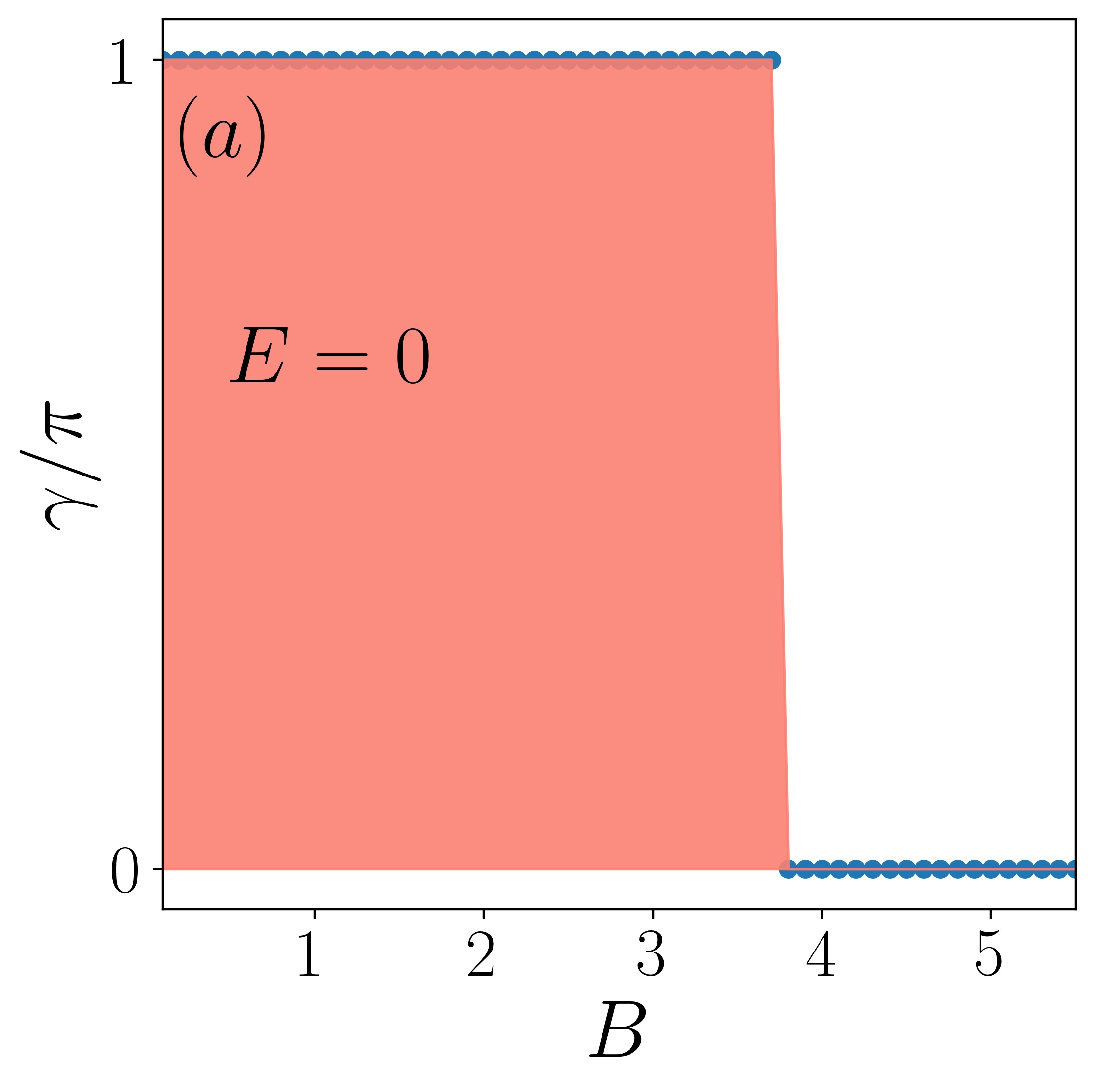}
\hfill
\hfill
\includegraphics[width=0.25\textwidth]{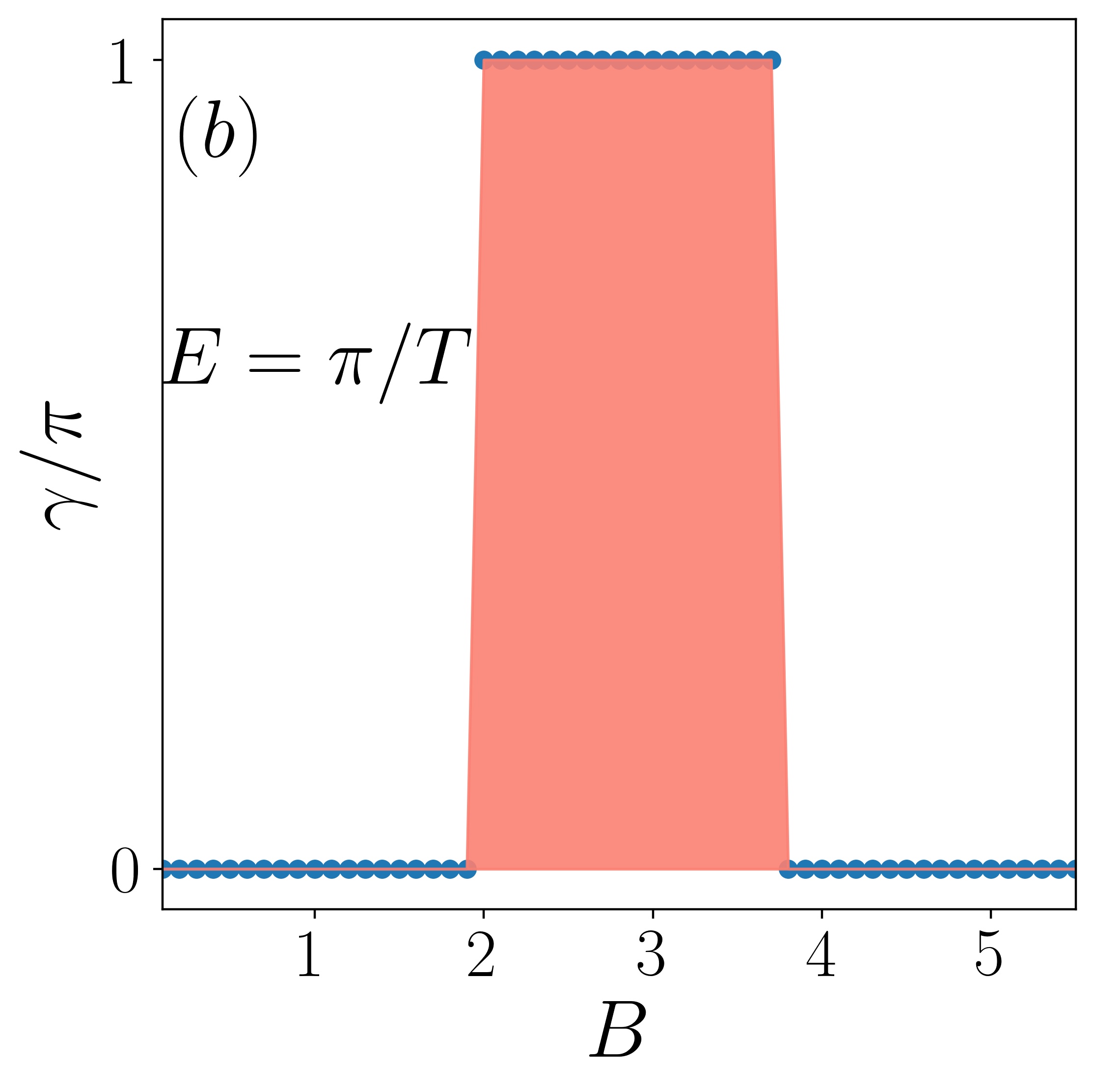}
\hfill}
\caption{The figures depict total Berry phases for the states filled upto $E = 0$ (panel a) and $E=\pi/T$ (panel b), as a function of the driving frequency, $\omega$ in the static trivial limit, $B>B_{c_2}$. The results correctly match with the real space quasi-energy spectrum shown in Fig.~\ref{Fig3}.}
\label{Fig5}
\end{figure}
\begin{subequations}
\begin{align}
    \omega_{c_1} &= -B + \sqrt{(\mu - 4t_{hop})^2 + \Delta^2} \\
    \omega_{c_2} &= B + \sqrt{(\mu - 4t_{hop})^2 + \Delta^2} \\
    \omega_t &= -B + \sqrt{\mu^2 + \Delta^2}
\end{align}
\end{subequations}
The gap closes at the center of the FBZ ($k=0$) corresponding to $\omega = \omega_t$, and at the edges of the FBZ ($k= \pm \pi$) for $\omega = \omega_{c_1,c_2}$. These conditions for the closure of the bulk gap serve to delineate the boundaries of the topological phases, as illustrated in Fig.~\ref{Fig3}. For instance, in Fig.~\ref{Fig3}a, for $B=0.5$, within the static trivial limit ($B<B_{c_1}$), the zero-energy modes persist up to $\omega = \omega_{c_2} = 4.5$. It is important to note that there are instances of gap closing, although not directly linked to the topological phase transition, however can result in the emergence of multiple Majorana modes. For example, in Fig.~\ref{Fig3}a, the MZMs experience an additional gap-closing transition at $\omega = \omega_{c_1}=3.5$ followed by another transition at $\omega = \omega_{t}=0.5$, both of which does not impact its topological phase. On the other hand, for $B=1.5$ (Fig.~\ref{Fig3}b), corresponding to static topological limit ($B_{c_1} < B < B_{c_2}$), MZMs appear beyond $\omega = \omega_{c_1}=2.5$ and undergo another gap closing transition at $\omega = \omega_{c_2}=5.5$, indicating an increase in the number of Majorana modes. Additionally, there are instances of gap closing for example, in Fig.~\ref{Fig3}b  $\omega \approx 1$, corresponds to a bulk gap closure at some non-high symmetric points ($k\neq 0, \pm \pi$). Finally, in Fig.~\ref{Fig3}c for $B=5$, corresponding to static trivial limit ($B> B_{c_2}$), the MZMs persist upto a frequency, $\omega = \omega_{t}$. However, in this scenario, there occurs only one gap-closing transition. Hence, based on the preceding discussions, we can infer that in the static non-trivial range, $B_{c_1} < B < B_{c_2}$, the driven system exhibits MZMs in the high-frequency regime ($\omega > \omega_{c_1}$). Whereas, in the static trivial range for $B< B_{c_1}$ and for $B>B_{c_2}$, the driven system showcases MZMs in the low-frequency regime, namely,  $\omega<\omega_{c_2}$ and $\omega<\omega_{t}$ respectively but not in the high-frequency regime.
\begin{figure}[t]
    \begin{subfigure}[b]{\columnwidth}
         \includegraphics[width=\columnwidth]{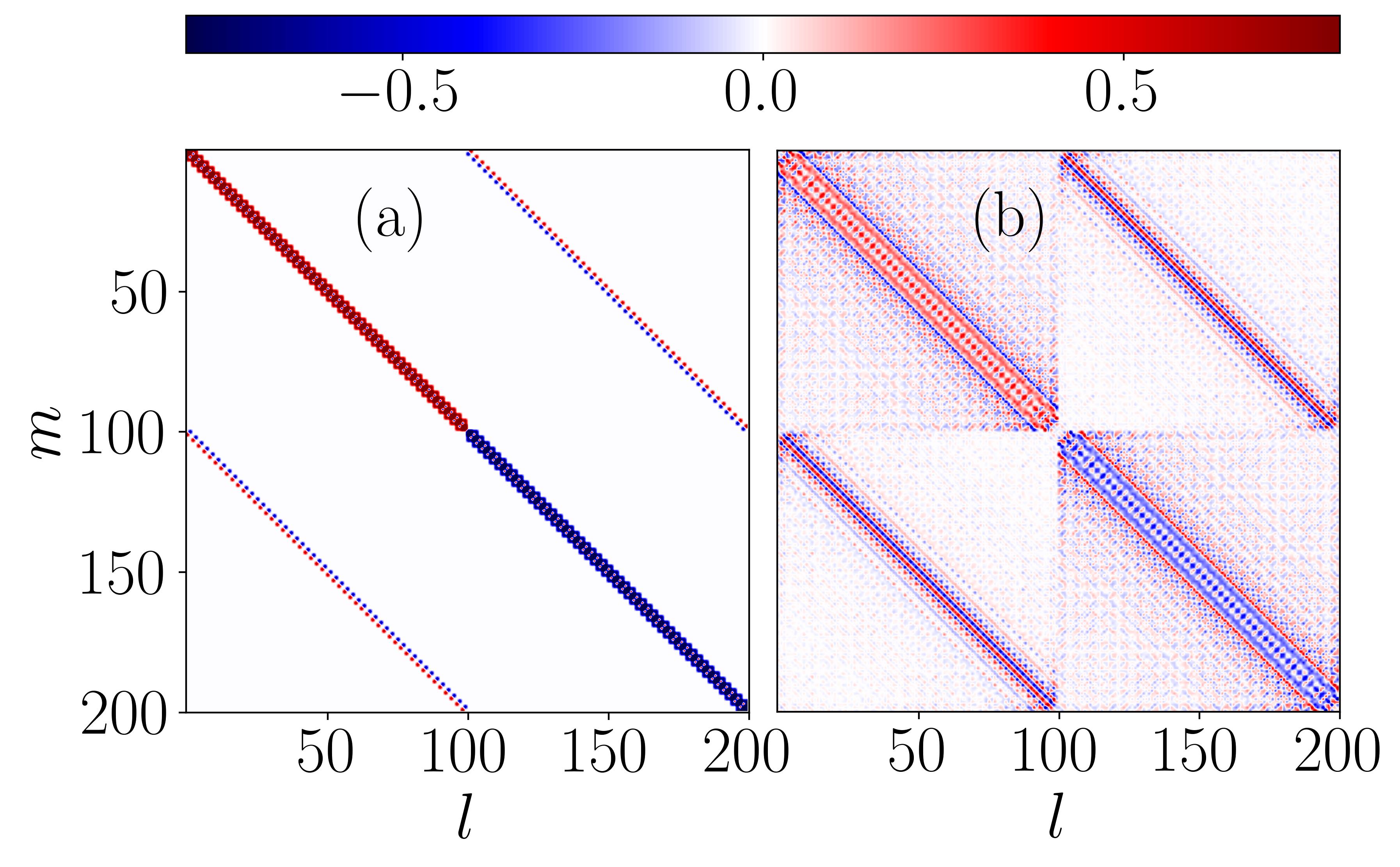}
        %  \caption{}
         \label{Fig2}
     \end{subfigure}
\caption{{Figure depicts the expansion coefficients for both static (panel a) and driven (panel b) Hamiltonian in the coordinate basis ($c_{1, \uparrow}^{\dagger}, c_{1, \downarrow}^{\dagger}, ... c_{N, \downarrow}^{\dagger}, c_{1, \uparrow}, c_{1, \downarrow}, ...c_{N, \downarrow}$), where $l$ and $m$ denotes the base indices. The results clearly indicates the induction of longer-range interactions. The rest of the parameters are chosen as, $B_1=1$, $B=1.5$, and $\omega=2$.}} 
\label{Fig6}
\end{figure}
\par To validate the topological characteristics alongside the `\emph{bulk-edge correspondence}', we compute the topological invariant \cite{10fold}. In the frequency domain, the pertinent invariant for a $3\times3$ Floquet-Bloch Hamiltonian is the Berry phase \cite{Berryphase,berryphase1,berryphase2}, that denotes the geometric phase accrued by a wave function during a smooth traversal across the Brillouin zone. The Berry phase is defined as,
\begin{equation}
\gamma=i\oint dk \langle{u_k}| \nabla_{k}{u_k}\rangle,
\end{equation}
where $\ket{u_k}$ are the Bloch states. A Hamiltonian exhibiting a non-trivial Berry phase cannot be adiabatically linked to an atomic insulator unless a gap-closing transition takes place. The numerical calculation of the Berry phases $\gamma_\beta$ corresponding to different bands ($\beta$ denotes the band index, marked with the letters $A-L$ in Fig.~\ref{Fig4}) for a particular frequency, say, $\omega=4$, is obtained as,
\begin{equation}
\gamma_{\beta}=
\begin{cases}
0 \quad (\beta= B,C,D,E,H,I,J,K) \\
\pi \quad (\beta= F,G)\\
\end{cases}
\textrm{for} 
 ~B < B_{c_1},
\end{equation}
\begin{equation}
\gamma_{\beta}=
\begin{cases}
0 \quad (\beta= F,G) \\

\pi \quad (\beta=B,C,D,E,H,I,J,K)\\
\end{cases}
\textrm{for} 
 ~ B_{c_1}<B<B_{c_2}.
\end{equation}
The outermost bands, that is, $\beta= A~\text{and}~L$ always contribute to zero Berry phase, since they do not participate in the band inversion process. It is noteworthy that there is at least one band below the Fermi level exhibiting a non-zero Berry phase, indicating that the system is in a topologically non-trivial state. One can also verify that the combined sum of Berry phases given by, $\gamma=|\textrm{mod}(\sum_{\beta} \gamma_\beta,2)|$, below an energy zero or $\pm \pi/T$ energy correlates with the edge modes identified in the real-space spectrum. For example, Fig.~\ref{Fig5} illustrates the cumulative sum of Berry phases up to $E=0$ (Fig.~\ref{Fig5}a) and $E=\pi/T$ (Fig.~\ref{Fig5}b) respectively corresponding to one of the static trivial conditions, namely, $B>B_{c_2}$. These correctly anticipate the appearance of zero and $\pi$ energy modes as already depicted in Fig.~\ref{Fig3}c.
\begin{figure*}[t] % Use figure* to span both columns
\begin{minipage}{\linewidth}
        \begin{subfigure}[t]{0.32\columnwidth}
            \includegraphics[width=\linewidth]{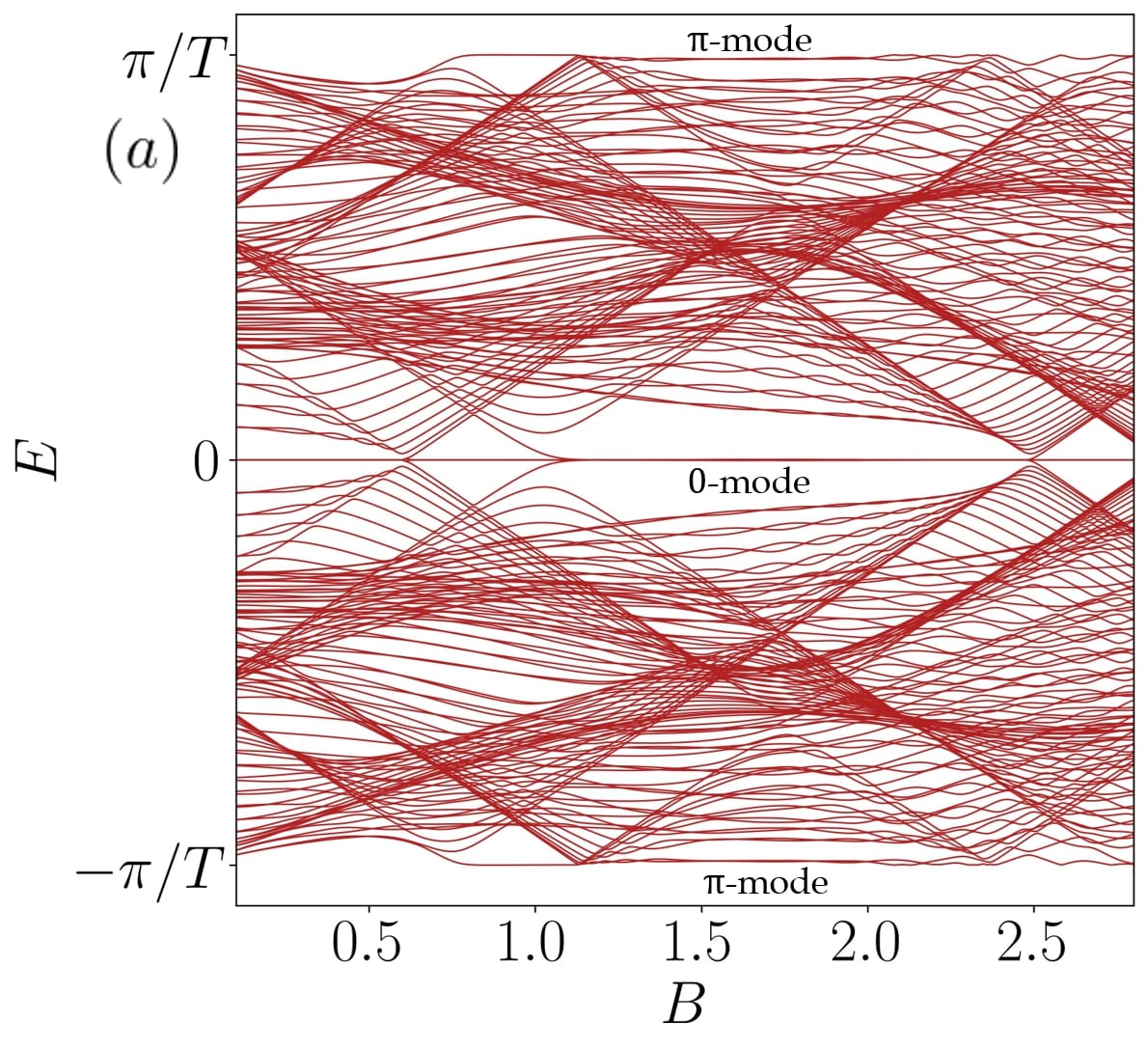}
            %\caption{}
        \end{subfigure}%
        \hfill
        \begin{subfigure}[t]{0.33\columnwidth}
            \includegraphics[width=\linewidth]{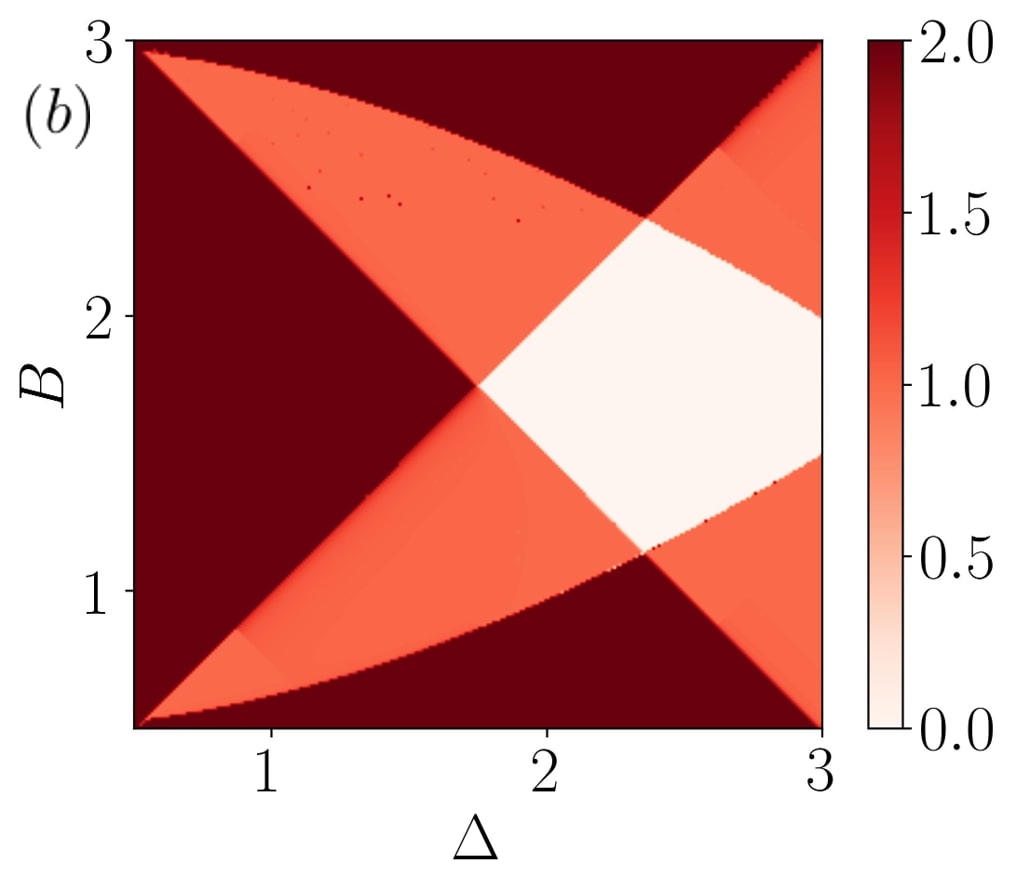}
            %\caption{}
        \end{subfigure}%
        \hfill
        \begin{subfigure}[t]{0.33\columnwidth}
            \includegraphics[width=\linewidth]{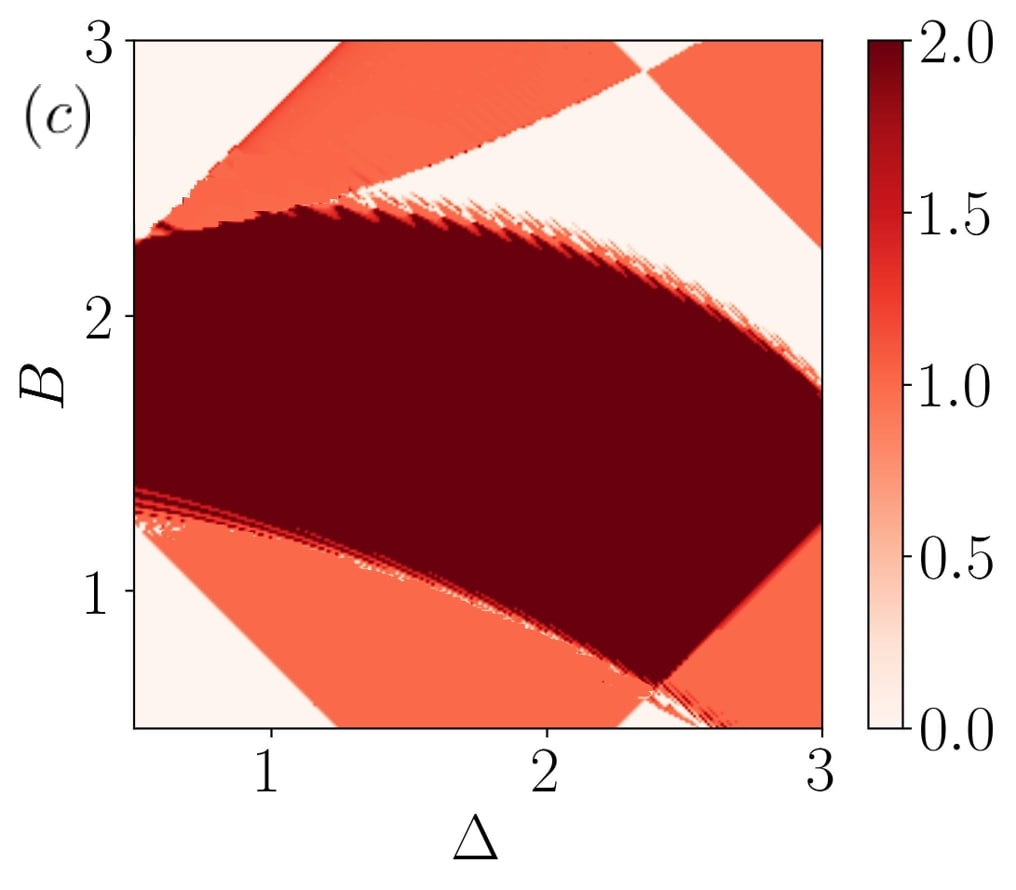}
            %\caption{}
        \end{subfigure}
\caption{Panel (a) shows the Floquet quasi-energy spectrum confined within the FBZ, plotted as a function of the strength of the magnetic field, $B$ with a fixed pairing strength, $\Delta=1$. Panel (b)-(c) depict the topological phase diagram in the $B-\Delta$ plane, computed using the winding numbers corresponding to the zero $(\nu^0)$ and the $\pi$ ($\nu^{\pi}$) energy modes respectively. The rest of the parameters are chosen as, $B_1=1,\omega=3.5$.}
\label{Fig7}
    \end{minipage}
\end{figure*}
\par Though the Berry phase provides insights into the emergence of Majorana zero and $\pi$ energy modes, it cannot precisely quantify the number of each of these localized edge modes. Consequently, we turn to calculating the particular invariant associated with the static version of the model. The static model as stated earlier falls in the symmetry class D which is characterized by a $\mathbb{Z}_2$ invariant, which can be computed either via the Pfaffian invariant or Fermion parity of the Majorana modes etc. Thus, one should obtain at most one pair of Majorana modes. The scenario raises doubts about the multiple gap-closing transitions observed and the prediction of many Majorana modes, as obtained by us here. This prompts a speculation regarding its origin. Could it be the emergence of longer-range interactions alongside the restoration of TRS in the driven system? This leads to one of the key results and shows the richness of the non-equilibrium phenomena embedded in our model. In order to validate our concern, we focus on diagonalizing the stroboscopic time evolution operator, $\hat{U}(T)$ introduced in Eq.~\ref{evolutionoperator}. In Fig.~\ref{Fig6}, we present the expansion coefficients of $H_{\text{eff}}$ in the coordinate basis ($c_{1, \uparrow}^{\dagger}, c_{1, \downarrow}^{\dagger}, ... c_{N, \downarrow}^{\dagger}, c_{1, \uparrow}, c_{1, \downarrow}, ...c_{N, \downarrow}$) corresponding to a particular frequency, namely, $\omega=2$, in order to compare and contrast it with the expansion coefficient for the static scenario illustrated in Fig.~\ref{Fig6}a. The analysis indicates a clear distinction, the coefficients in the driven scenario reveal evidence of longer-range (spanning over
% \begin{widetext}
%     \begin{minipage}{\linewidth}
%         \begin{figure}[H]
%         %\centering
%         \hspace{-12mm}
%           \begin{subfigure}[!h]{0.33\columnwidth}
%          %\centering
%          \includegraphics[width=\columnwidth]{Figure_7a.jpg}
%          %\caption{}
%          \label{Fig7.1}
%      \end{subfigure}
%      \begin{subfigure}[!h]{0.34\columnwidth}
%          \includegraphics[width=\columnwidth]{Figure_7b.jpg}
%          %\caption{}
%          \label{Fig7.2}
%      \end{subfigure}
%      \begin{subfigure}[!h]{0.34\columnwidth}
%          \includegraphics[width=\columnwidth]{Figure_7c.jpg}
%          %\caption{}
%          \label{Fig7.3}
%      \end{subfigure}
% \caption{Panel (a) shows the Floquet quasi-energy spectrum confined within the FBZ, plotted as a function of the strength of the magnetic field, $B$ with a fixed pairing strength, $\Delta=1$. Panel (b)-(c) depict the topological phase diagram in the $B-\Delta$ plane, computed using the winding numbers corresponding to the zero $(\nu^0)$ and the $\pi$ ($\nu^{\pi}$) energy modes respectively. The rest of the parameters are chosen as, $B_1=1,\omega=3.5$.} 
%  \label{Fig7}
% \end{figure}    
% \end{minipage}
% \end{widetext}
more than 10 sites) interactions (Fig.~\ref{Fig6}b). Moreover, as the frequency decreases, more longer-range interactions emerge in the driven system. The way one can reconcile this is to do a Baker-Campbell-Hausdorff (BCH) expansion where the nested commutator brackets induce coupling between distant sites which yields more off-diagonal terms in $\hat{U}(T)$ leading to the generation of many Majorana modes. Subsequently, we shall address this issue by leveraging all the symmetries inherent in the evolution operator, $\hat{U}(T)$. 
\par In a generic sense, Floquet systems break the original symmetries associated with the static version of the model such as TRS and Chiral symmetry (CS) \cite{broken_symmetry,broken_symmetry2,chiral_restoration2}. Interestingly, different choices of the time period can result in the effective Hamiltonian ($H_{\text{eff}}$) possessing different symmetries or none at all. To ensure, the necessary symmetries of the system, there has to be an intermediate time, $t$ ($t_0<t< t_0 + T$) that splits the period into two symmetrical parts \cite{asbothwinding1,asbothwinding2}. Let $F$ and $G$ denote the time evolution of the first and second part of the cycle respectively, that is,
\begin{equation}
    F = \mathcal{T} e^{-i \int_{t_0}^{t} H(t) dt} ~ ; ~ G = \mathcal{T} e^{-i \int_{t}^{t_0 + T} H(t) dt}. 
\end{equation}
Now, suppose these two equivalent parts of the cycle are chiral symmetric to each other (that is, $S F^{\dagger} S^{-1} = G$). In that case, one can perform a similarity transformation $F \hat{U}(T) F^{-1}$ or equivalently, $G^{-1} \hat{U}(T) G$ to construct the time evolution operators corresponding to two \textit{symmetric time frames} which further can be represented as, $U_1 = FG$ and $U_2 = GF$ respectively \cite{chiral_restoration1,chiral_restoration2}. These operators not only retain the same quasi-energy spectrum as the original Floquet Hamiltonian ($H_{\text{eff}}$) but also preserve all the essential symmetries of the static counterpart that were lost in the driven scenario. Additionally, these representations allow us to reconstruct several symmetry operations, such as TRS and CS via certain new operators\cite{chiral_restoration1}. For instances,
\begin{equation}
    \text{TRS:} ~ \hat{F}^{-1}(-k) \hat{\kappa} \hat{F}(k) \hat{U}(T,k) \hat{F}^{-1}(k) \hat{\kappa} \hat{F}(-k) = \hat{U}^{\dagger}(T,-k), 
\end{equation}
and
\begin{equation}
    \text{CS:} ~ \hat{F}^{-1}(k) \hat{S} \hat{F}(k) \hat{U}(T,k) \hat{F}^{-1}(k) \hat{S}^{-1} \hat{F}(k) = \hat{U}^{-1}(T,k), 
\end{equation}
where, $\hat{S} = \tau_x \sigma_0$. It implies, that the driven Hamiltonian can possess a hidden symmetries under the newly developed TRS and CS operator. Equivalently, $\hat{T} = \hat{G}(-k) \hat{\kappa} \hat{G}^{-1}(k)$ and $\hat{S}^{\prime} = \hat{G}(k) \hat{S} \hat{G}^{-1}(k)$ plays the similar role as TRS and CS operators respectively. By carefully selecting the period of the drive in order to preserve all three symmetries, namely particle-hole symmetry, as discussed in Sec. ~\ref{sec:level2} along with the newly formed TRS and CS operators, the driven system can be characterized via a pair of $\mathbb{Z} \times \mathbb{Z}$ invariant. Furthermore, in the context of 1D systems with restored CS, the Hamiltonian is effectively classified in the first homotopy group \cite{10fold,schnyder}, providing us with the winding number as the topological invariant that is going to count the total number of localized edge states. The conventional definition of the winding number is given as,
\begin{equation}
    \nu[h]=\frac{1}{2\pi i}\int_{BZ} dk \frac{d}{dk} \ln {h(k)}.
    \label{Eq:winding}
\end{equation}
Mathematically, the winding number assumes the same information as the Berry phase. However, determining the winding number entails the Hamiltonian to adopt an off-diagonal form in the canonical basis, where $\hat{S}$ is diagonal. This, for our $\mathcal{H}(k)$ reads as,
\begin{equation}
    \hat{R}^\dagger \mathcal{\hat{H}}(k) \hat{R} = \begin{pmatrix} 0 & A (k)\\ A^{\dagger} (k) & 0\\
\end{pmatrix},
\end{equation}
where $A(k)$ is a $2 \times 2$ Hermitian matrix. Further, $\hat{R}$ is a unitary matrix constructed using the chiral basis,
\begin{equation}
    \hat{R} =  \frac{1}{\sqrt{2}} \begin{pmatrix} \sigma_x & \sigma_x \\
-\sigma_y & \sigma_y\\
\end{pmatrix},
\end{equation}
Since, $\det(\hat{R}^\dagger \mathcal{\hat{H}}(k) \hat{R}) = det(A(k)) det(A^{\dagger}(k))$, $det(A(k))$ can only vanish if $\mathcal{H}(k)$ has a vanishing determinant or equivalently a zero eigenvalue. Therefore, the Hamiltonian, $\mathcal{H}(k)$, with a gap at zero-energy can be characterized by a complex function, $z(k) = e^{i\theta(k)} = det(A(k)) / |det(A(k))|$. Consequently, one can replace $h(k)$ in Eq.~\ref{Eq:winding} by $h(k) = det(A(k))$ to get the expression for the winding number as,
\begin{equation}
   \nu = \frac{1}{2\pi i} \oint dk \frac{d}{dk} \ln{det|A|}.
\end{equation}
% The symmetry classification of $H_{\text{eff}}$ aids in classifying $\hat{U}(T,0)$. However, it turns out that $H_{\text{eff}}$ may have various symmetries or none at all, depending on the choices of the period selection made for $\hat{U}(T,0)$. 
However, the invariants associated with either of the frames ($\hat{U}_1$ or $\hat{U}_2$) can not individually offer complete insight into the number of edge modes. Rather, based on the periodic table of Floquet topological insulators and superconductors \cite{roy10fold}, they need to be combined in a specific way so that each non-trivial phase can be identified by a pair of noncommutative winding numbers given as,
\begin{equation}
    \nu^0 = \frac{\nu^{\prime} + \nu^{\prime \prime}}{2} \quad ; \quad \nu^{\pi} = \frac{\nu^{\prime} - \nu^{\prime \prime}}{2}.
\end{equation}
Here $\nu^{\prime}$ and $\nu^{\prime \prime}$ are the winding numbers for the two effective Hamiltonians corresponding to the two symmetric time frames $\hat{U}_1$ and $\hat{U}_2$ respectively. Consequently, the emergence of longer-range interactions alongside the restoration of TRS allows us to have higher winding numbers, leading to the appearance of multiple Majorana modes.
\par Fig.~\ref{Fig7} shows the topological phase diagrams in $B-\Delta$ plane, plotted for certain values of $\omega$ and $B_{1}$ such that, $\omega= 3.5$, and $B_{1}=1$. To confirm the \emph{bulk-edge correspondence}, one can compare the outcomes with the real-space quasi-energy spectrum plotted as a function of $B$ for $\Delta = 1$ (Fig.~\ref{Fig7}a). A noteworthy observation from Figs.~\ref{Fig7}a,b is the coexistence of zero and $\pi$ Majorana modes in certain regions of the parameter space which can lead to intriguing results as discussed in recent investigations such as, \cite{period2t1,period2t2}.
%\vspace{0.5in}
% \begin{center}{\subsection{\label{sec:level3.1}Results}}\end{center}
\vspace{0.3in}
\begin{center}{\section{\label{sec:level4}Multi-frequency drive}}\end{center}
\vspace{-0.3in}
\begin{center}{\subsection{\label{sec:level3.1}Formalism}}\end{center}
\textit{Commensurate case:} Here, we present the method for the treatment of Eq.~\ref{Ham_zeeman_periodic} corresponding to a two-tone commensurate driving setup with $B_2 \neq 0$. The periodicity of the two drives, $T_1$ and $T_2$ may obey following equation,
\begin{equation}
    \frac{T_1}{T_2} = \frac{n_2}{n_1},
\end{equation}
implying that one may always find a common time period $T$, such that,
\begin{equation}
    T = n_1 T_1 = n_2 T_2
\end{equation}
% \begin{figure}[t]
%           \begin{subfigure}[!h]{0.493\columnwidth}
%          %\centering
%          \includegraphics[width=\columnwidth]{Picture12.png}
%          \captionlistentry{}
%          \label{8.1}
%      \end{subfigure}
%      \begin{subfigure}[!h]{0.493\columnwidth}
%          \includegraphics[width=\columnwidth]{Picture13.png}
%          \captionlistentry{}
%          \label{8.2}
%      \end{subfigure}
%      \begin{subfigure}[!h]{0.493\columnwidth}
%          \includegraphics[width=\columnwidth]{Picture14.png}
%          \captionlistentry{}
%          \label{8.3}
%      \end{subfigure}
%      \begin{subfigure}[!h]{0.493\columnwidth}
%          \includegraphics[width=\columnwidth]{Picture15.png}
%          \captionlistentry{}
%          \label{8.4}
%      \end{subfigure}
% \caption{{The figure depicts topological phase diagrams concerning MZMs for single-drive scenario (panel a) and for various multi-frequency driving scenario. For example, we study the case of a 1:2 drive in panel (b), 2:3 drive in panel (c), and a 4:5 drive in panel (d). All the different driving scheme leads to the same gap closure and the same winding number in the region ($B,T$)$\le (1.5,1.5)$. the strength of the two drives are chosen as, $B_1 = B_2 = 1$.}} 
%  \label{Fig8}
% \end{figure}
which will further be used to employ the Floquet formalism as discussed in Sec.~\ref{sec:level2}.
\newline
\textit{Incommensurate case:} In a similar fashion, one can explore the implications on the topological characteristics when the Rashba nanowire model with topological superconductivity is subjected to an incommensurate multi-frequency driving protocol, where the ratio $n_1:n_2$ is no longer a rational number. For instance, let us consider the case where $n_1:n_2 = \beta$, where $\beta = (\sqrt{5}+1)/2$, represents the \emph{golden ratio}, defined as the ratio between two successive large numbers in the Fibonacci series. Consequently, with no global time periodicity, the conventional stroboscopic Floquet operator cannot be defined. To address this, we shall utilize the framework established by Ho et al. in Ref.~\cite{multimode1,multimode2,multimode3,validityofmultimode}, also referred to as the many-mode Floquet theory (MMFT). In a generic sense, MMFT is essentially an extension of the standard Shirley-Floquet (SF) formalism to scenarios involving more than one frequency. By utilizing the Shirley-Floquet approach as a foundation, one can set up the formalism by relabelling the basis vectors in the Shirley's framework to construct an extended basis vector derived from the tensor product of the two Fourier spaces. That is,
\begin{equation}
 \ket{m} \rightarrow \ket{m_1} \otimes \ket {m_2},
\end{equation}
with the initial condition being, $m\omega = m_1 \omega_1 + m_2 \omega_2$ or equivalently, $m = n_1m_1 + n_2m_2$, where $m_1,m_2$ are two integers. representing the Hilbert space of time-periodic functions spanned by $e^{-im_{(1,2)} \omega t}$.
Consequently, without loss of generality, based on the Shirley-Floquet formalism elucidated in sec.~\ref{sec:level2}, the multi-frequency time-dependent problem can be mapped to an equivalent time-independent eigenvalue problem (similar to Eq.~\ref{Eq:SF3}) given as,
\begin{equation}
\begin{split}
    \bra{m_{1},m_{2}} H_F \ket{m_{1}^{\prime},m_{2}^{\prime}}  = & (m_1 \omega_1 + m_2 \omega_2) \delta_{m_{1},m_{1}^{\prime}} \delta_{m_{2},m_{2}^{\prime}}  
    \\ & +  E_0 \delta_{m_1,m_{1}^{\prime}}
    \delta_{m_{2},m_{2}^{\prime}} \\ &+  \mathcal{H}^{(m_{1}-m_{1}^{\prime}),(m_{2}-m_{2}^{\prime})},
\end{split}
\end{equation}
where $E_0$ is the eigenvalue corresponding to the static part of the Hamiltonian, $H_0$, and the off-diagonal terms, $\mathcal{H}^{(m_{1}-m_{1}^{\prime}),(m_{2}-m_{2}^{\prime})}$ are defined as,
\begin{equation}
\begin{split}
    \mathcal{H}^{(m_{1}-m_{1}^{\prime}),(m_{2}-m_{2}^{\prime})}  = & \sum_{i=1}^2 B_{i} /2 \sum_j  \Big [\hat{c}^\dagger _{j,\uparrow} \hat{c}_{j,\downarrow} + \hat{c}^\dagger _{j, \downarrow} \hat{c} _{j,\uparrow} \Big ]  \\ & \times \Big( \delta_{m_{i}-m_{i}^{\prime},1} + \delta_{m_{i}-m_{i}^{\prime},-1}\Big).
\end{split}
\label{Eq:MMFT1}
\end{equation}
In matrix notation, $H_F$ read as,
\begin{widetext}
    \begin{minipage}{\linewidth}
    \begin{equation}
%\hspace*{-0.7cm}
    H_{F} = {\begin{bmatrix}\ddots & \vdots  & \vdots & \vdots &\vdots & \vdots & \iddots \\ \dots & A - 2\omega_2 & B & 0 & 0 & 0 & \dots \\ \dots & B & A  - \omega_2 & B & 0 & 0 & \dots \\ \dots & 0 & B & A & B & 0 & \dots \\ \dots & 0 & 0 & B & A +\omega_2 & B & \dots \\
    \dots & 0 & 0 & 0 & B & A + 2\omega_2 & \dots \\ \iddots & \vdots & \vdots & \vdots & \vdots & \vdots & \ddots
    \end{bmatrix}},
    \label{MMFT_compact_matrix}
\end{equation}
where,
    \begin{equation}
%\hspace*{-0.7cm}
    A = {\begin{bmatrix}\ddots & \vdots  & \vdots & \vdots &\vdots & \vdots & \iddots \\ \dots & H_0 - 2\omega_1 & V_1 & 0 & 0 & 0 & \dots \\ \dots & V_1 & H_0  - \omega_1 & V_1 & 0 & 0 & \dots \\ \dots & 0 & V_1 & H_0 & V_1 & 0 & \dots \\ \dots & 0 & 0 & V_1 & H_0 +\omega_1 & V_1 & \dots \\
    \dots & 0 & 0 & 0 & V_1 & H_0 + 2\omega_1 & \dots \\ \iddots & \vdots & \vdots & \vdots & \vdots & \vdots & \ddots
    \end{bmatrix}}, \quad \text{and} \quad
    B = {\begin{bmatrix}\ddots & \vdots  & \vdots & \vdots &\vdots & \vdots & \iddots \\ \dots & V_2 & 0 & 0 & 0 & 0 & \dots \\ \dots & 0 & V_2 & 0 & 0 & 0 & \dots \\ \dots & 0 & 0 & V_2 & 0 & 0 & \dots \\ \dots & 0 & 0 & 0 & V_2 & 0 & \dots \\
    \dots & 0 & 0 & 0 & 0 & V_2 & \dots \\ \iddots & \vdots & \vdots & \vdots & \vdots & \vdots & \ddots
    \end{bmatrix}}
    \label{Eq:MMFT_matrix}
\end{equation}
    \end{minipage}
\end{widetext}
with $V_{i} = B_{i} /2 \sum_j \Big [\hat{c}^\dagger _{j,\uparrow} \hat{c}_{j,\downarrow} + \hat{c}^\dagger _{j, \downarrow} \hat{c} _{j,\uparrow} \Big ]$. Additionally, it is important to note that the matrix $A$ is nothing but the SF Hamiltonian corresponding to the single drive scenario, as shown in Eq.~\ref{Eq:SF2}.
\begin{figure}[t]
    \begin{subfigure}[b]{0.75\columnwidth}
         \includegraphics[width=\columnwidth]{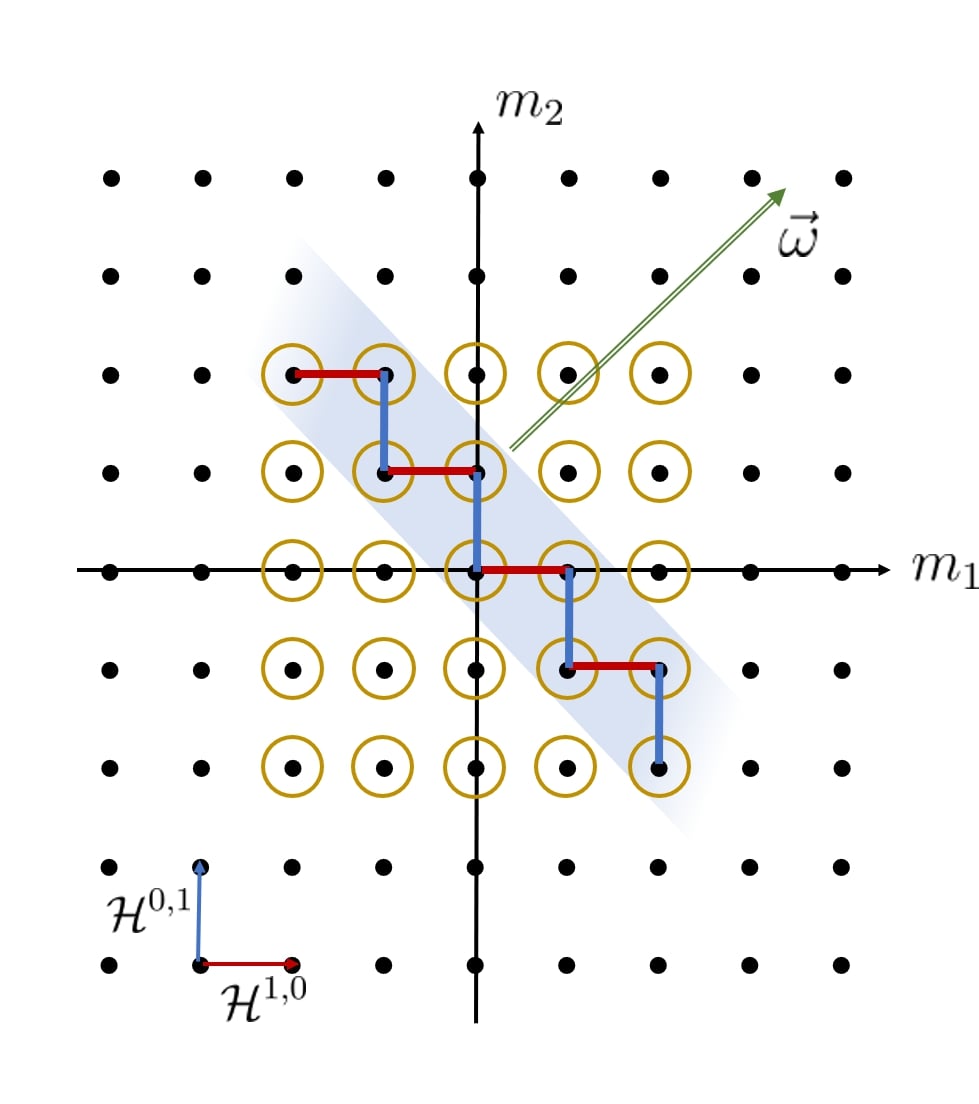}
        %  \caption{}
         \label{Fig1}
     \end{subfigure}
\caption{{Visualization of the many-mode Floquet theory using two-dimensional tight-binding representation spanned by the coordinates $m_1$ and $m_2$. The light blue strip denotes a quasiperiodic ladder, inside of which the nearest neighbor couplings are represented by either vertical hopping, $\mathcal{H}^{0,1}$ (blue solid line) or horizontal hopping, $\mathcal{H}^{1,0}$ (red solid line). In principle, the basis sets for MMFT calculations should run over all $m_1$ and $m_2$. However, a finite basis sets (consisting of lattice points in $m_1$-$m_2$ plane marked with circle) have been used to numerically diagonalize the MMFT Hamiltonian, shown in Eq.~\ref{Eq:MMFT1}}}. 
\label{Fig8}
\end{figure}
\par Similar to the SF formalism, the quasi-energy eigenvalue equation for a multi-driving protocol can be visualized as a tight-binding problem in a $d$-dimensional lattice, where the coordinates are denoted by $m_1,m_2,....m_d$, with $d$ being the number of drives. Additionally, there is a Stark field $\vec{\omega}$ pointing along the synthetic dimensions, effectively segmenting the entire lattice into layers of a quasiperiodic structure, represented by a blue strip (in case of two mutually irrational drives) as shown in Fig.~\ref{Fig8}. The quasiperiodic modulation arises since the on-site potential within this structure varies from site to site. For a given coordinate, the on-site energy is represented as $m_1 \omega_1 + m_2 \omega_2 + ...$. Consequently, the energy difference between two neighboring sites corresponds to one of the $\omega_{i}$ values, and since these $\omega_i$ values are irrational with respect to each other, the structure follows a quasiperiodic modulation. Additionally, By shifting the blue strip along the direction of the Stark field, the entire lattice can be generated. Thus, this quasiperiodic structure can be viewed as an admixture of $d$ Wannier-Stark ladders. Furthermore, corresponding to a two-tone harmonic drive, the ladder is confined to nearest-neighbor hopping, $\mathcal{H}^{m,m^{\prime}}$ (from $m^{\text{th}}$ to $(m+1)^{\text{th}}$ rung of the ladder), which can be either vertical, $\mathcal{H}^{0, 1}$ or horizontal, $\mathcal{H}^{1,0}$  (represented by blue and red solid lines, respectively, in Fig.~\ref{Fig8}), as mathematically defined by Eq.~\ref{Eq:MMFT1}. This conceptual framework provides a clear understanding of the interplay between multiple driving frequencies and their effects on the dynamics of the system. Furthermore, their interplay can lead to inequivalent particle-hole symmetric points at energies, $E=0, \omega_1/2$, and $\omega_2/2$ (see Fig.~\ref{Fig10}). Consequently, in an open chain, this leads to the emergence of distinct Majorana modes appearing other than zero energy that were absent in the static case. Thus, the localization of these unique Majorana modes, despite a dense spectrum, can be understood as the result of two independent localization mechanisms, namely, the Wannier-Stark localization in the direction of the Stark field, $\vec{\omega}$ and quasiperiodic localization within the drive-induced synthetic dimensions.
% \begin{figure}[t]
%     \begin{subfigure}[b]{\columnwidth}
%          \includegraphics[width=\columnwidth]{Figure_8.jpg}
%         %  \caption{}
%          \label{Fig1}
%      \end{subfigure}
% \caption{{The figure depicts topological phase diagrams concerning MZMs for single-drive scenario (panel a) and for various multi-frequency driving scenario. For example, we study the case of a 1:2 drive in panel (b), 2:3 drive in panel (c), and a 4:5 drive in panel (d). the strength of the two drives are chosen as, $B_1 = B_2 = 1$.}} 
% \label{Fig9}
% \end{figure}
\begin{figure}[t]
    \begin{subfigure}[b]{\columnwidth}
         \includegraphics[width=\columnwidth]{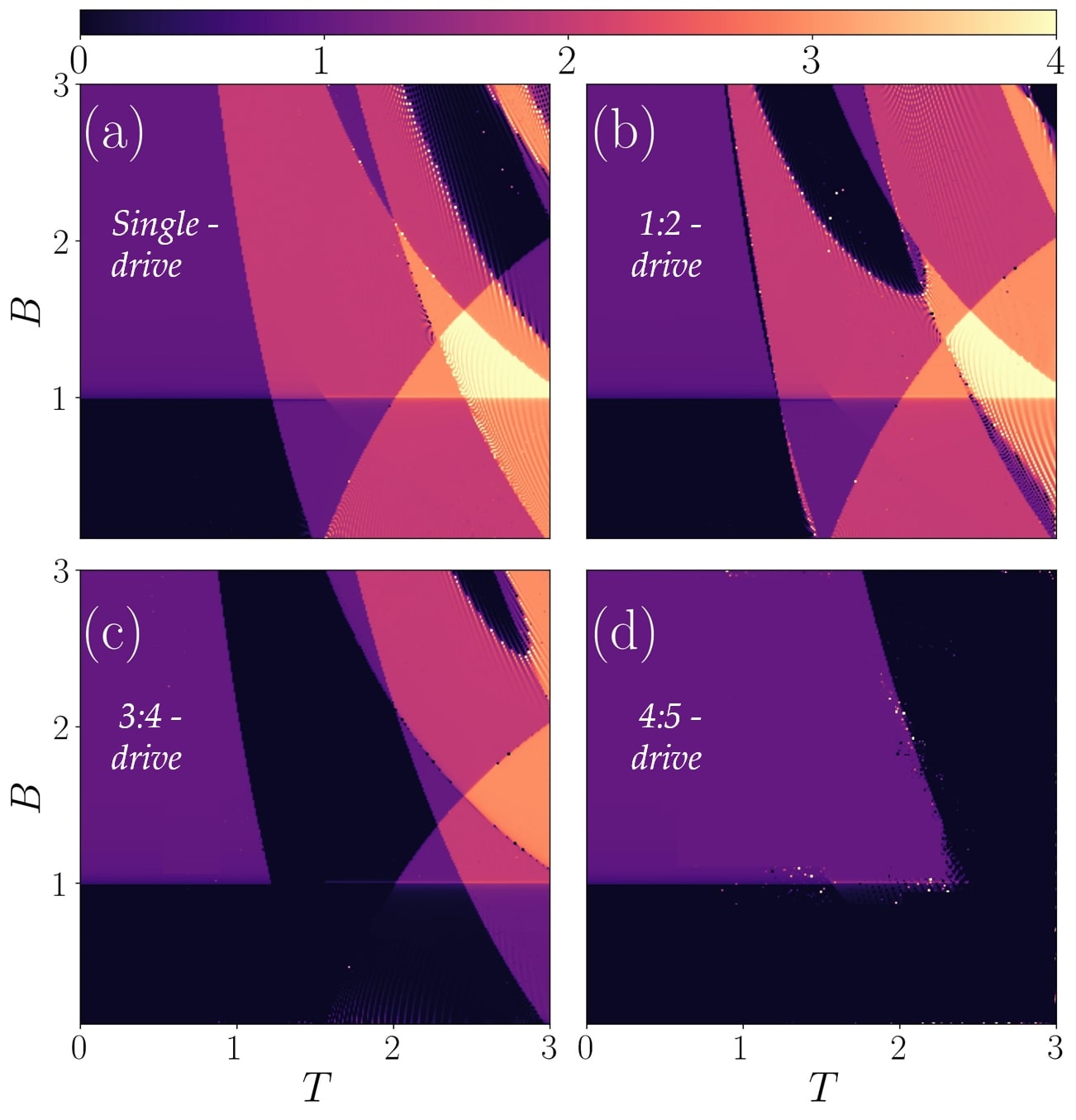}
        %  \caption{}  % Omit this line if not using subfigure captions
        %  \label{Fig1}  % Omit this line if not referring to the subfigure separately
     \end{subfigure}
\caption{{The figure depicts topological phase diagrams concerning MZMs for the single-drive scenario (panel a) and for various multi-frequency driving scenarios. For example, we study the case of a 1:2 drive in panel (b), a 2:3 drive in panel (c), and a 4:5 drive in panel (d). The strength of the two drives is chosen as $B_1 = B_2 = 1$.}} 
\label{Fig9}  % Place the label here to refer to the whole figure
\end{figure}
\par Moreover, the effective field $\vec{m} \cdot \vec{\omega}$ can indicate the absorption or emission of energy from the drive in terms of the integers, $m_1$ and $m_2$. Additionally, since $\vec{m} \cdot \vec{\omega} = m_1 \omega_1 + m_2 \omega_2$ can be seen to yield any arbitrary energy increment along both $m_1$ and $m_2$, the quasi-energy spectrum gets densely populated everywhere. Consequently, it appears that the Majoranas lack a protective cover to prevent them from hybridizing into the bulk. Nevertheless, these Majoranas remain stable and robust against local perturbations owing to the two independent localization schemes as discussed above.
\vspace{0.2in}
\begin{center}{\subsection{\label{sec:level3.2}Results}}\end{center}
% \begin{figure}[t]
%     \begin{subfigure}[b]{\columnwidth}
%          \includegraphics[width=\columnwidth]{Figure_8.jpg}
%         %  \caption{}
%          \label{Fig9}
%      \end{subfigure}
% \caption{{The figure depicts topological phase diagrams concerning MZMs for single-drive scenario (panel a) and for various multi-frequency driving scenario. For example, we study the case of a 1:2 drive in panel (b), 2:3 drive in panel (c), and a 4:5 drive in panel (d). the strength of the two drives are chosen as, $B_1 = B_2 = 1$.}} 
% \label{Fig9}
% \end{figure}
\textit{Commensurate case:} At first, we shall investigate the impact on the topological characteristics when the ratio of the two frequencies is commensurate and can be related as, $n_1:n_2$ = 1:2. This choice of bichromatic driving is considered experimentally feasible and has been demonstrated to possess adjustable sensitivity  \cite{rydberg}. 
\par It is worth noting that, maintaining the first drive constant, the additional drive results in minimal alterations to the nature associated with the $\pi$ modes, especially in the high-frequency region. Therefore, our subsequent discussion will concentrate on the evolution of zero modes under two-tone commensurate driving across different frequency regimes. In Fig.~\ref{Fig9}, we have illustrated topological phase diagrams showing the distribution of zero energy modes across the $B-T$ plane. To ensure a smooth comparison, the single-drive scenario has been included as well.  Upon analysis of these figures, notable findings emerge. For instance, regions that appeared topologically trivial in Fig.~\ref{Fig9}a now exhibit non-trivial behavior with the inclusion of an additional drive (Fig.~\ref{Fig9}b), and the converse is true as well. Furthermore, we have included results regarding other commensurate multi-tone driving protocols, where the ratios of integers are not uniquely expressed as an integer multiple, such as $n_1:n_2$ = 3:4 (Fig.~\ref{Fig9}c) and $n_1:n_2$ = 4:5 (Fig.~\ref{Fig9}d) to encompass more intricate driving scenarios. It is evident that such intricate driving patterns tend to populate the spectrum more densely, leading to quasi-energy spectra that are denser and thus reducing the bulk gap compared to that of the single-drive and 1:2 cases. Consequently, lesser number of localized edge modes appear to be shielded by a finite bulk gap, thereby shrinking the range of non-trivial phases. Moreover, even in Fig.~\ref{Fig9}d, corresponding to a 4:5 driving protocol, the system has no longer access to higher winding numbers. Nonetheless, regardless of the different driving protocols, they all lead to very similar gap closure and hence the same winding numbers are obtained in the $B-T$ plane, specifically in the region where $(B,T)\le(1.5,1.5)$. The alteration in the winding number only occurs at higher values of magnetic field, $B$, and the time period, $T$.
\par From the preceding discussions, it is clear that the localization of the edge modes is significantly influenced by the chosen driving protocol. Even with more intricate driving scenarios (where the integers are not uniquely expressed as integer multiples), the Majorana modes lose their characteristic localization behavior and manifest as extended states. Thus, it is conceivable to manipulate the stability and presence of the edge modes by conveniently switching between different driving protocols. This could prove beneficial in quantum computation setup comprising of multiple interconnected segments of Kitaev chains \cite{kitaevquantumcomputationfloquet}. The experimental advantage lies in the ability to selectively alter the localization of the edge modes without changing the static parameters of the model. As a result, one could apply this multi-tone driving strategy to a network of Kitaev chain segments, where the parameters of each segment may be beyond controllable due to practical limitations. It is worth mentioning that the bichromatic driving protocol can also be examined using the Shirley-Floquet approach, as previously discussed in Sec.~\ref{sec:level2}. However, truncating the infinite-dimensional matrix will now depend on the ratio of the driving frequencies. Consequently, corresponding to a complex driving scenario utilizing this formalism and performing associated computations could appear to be cumbersome \cite{solin}.
\begin{figure}[t]
    \begin{subfigure}[b]{0.75\columnwidth}
         \includegraphics[width=\columnwidth]{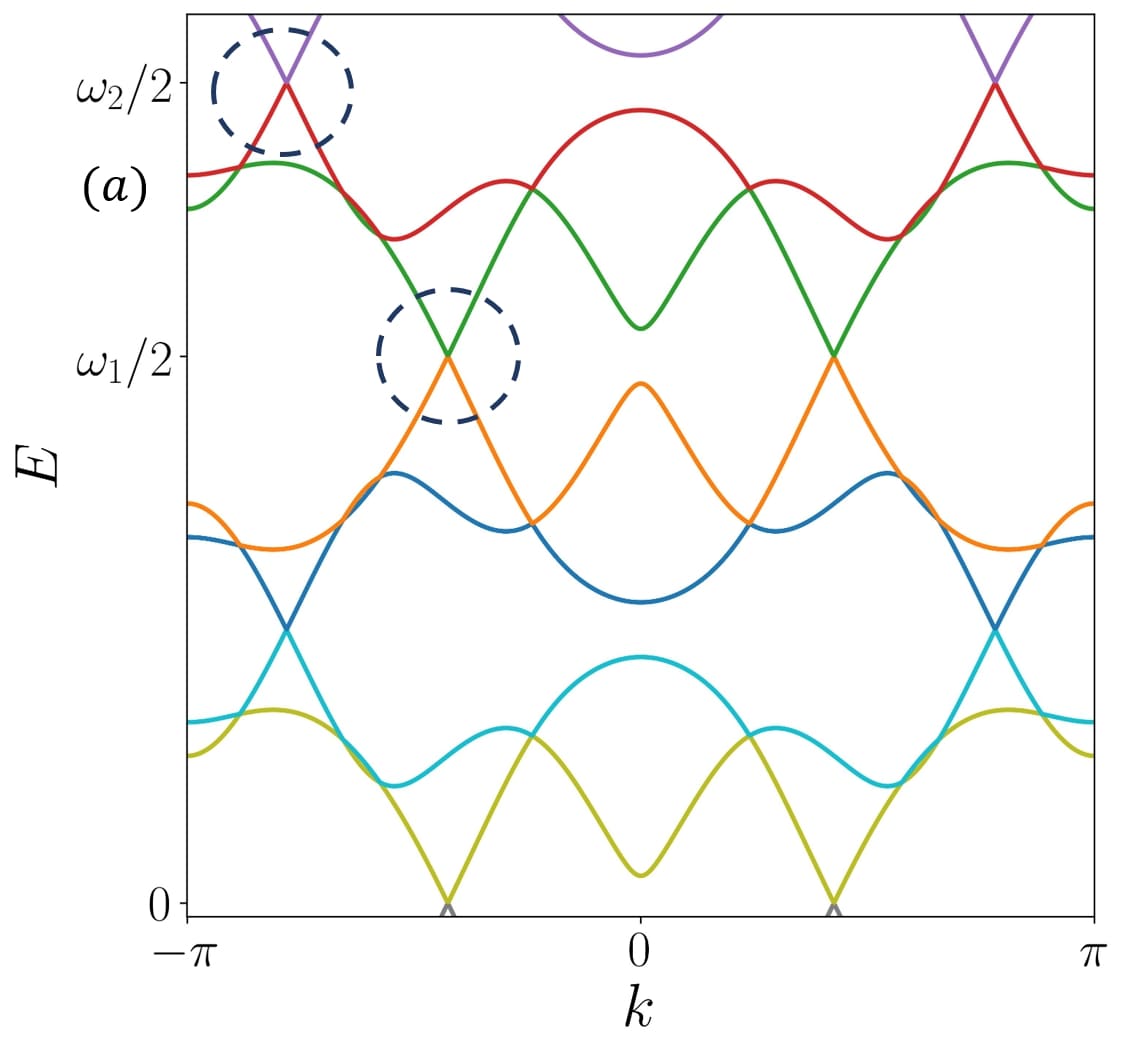}
        %  \caption{}
         \label{Fig1}
     \end{subfigure}
\caption{{The figure depicts the bulk quasi-energy spectrum corresponding to an incommensurate multi-frequency driving protocol in the static topological limit ($B=1.5$). There are two distinct level crossings at energy, $E=\omega_1 /2$, and $E=\omega_2 /2$ (as marked by dashed circle). The ratio of the two frequencies has been chosen as, $\omega_2/\omega_1 = \beta= (\sqrt{5}+1)/2.$ The strength of the two drives are fixed at, $B_1=B_2=1$.}} 
\label{Fig10}
\end{figure}
\begin{figure}[!t]
\centerline{\hfill
\includegraphics[width=0.25\textwidth]{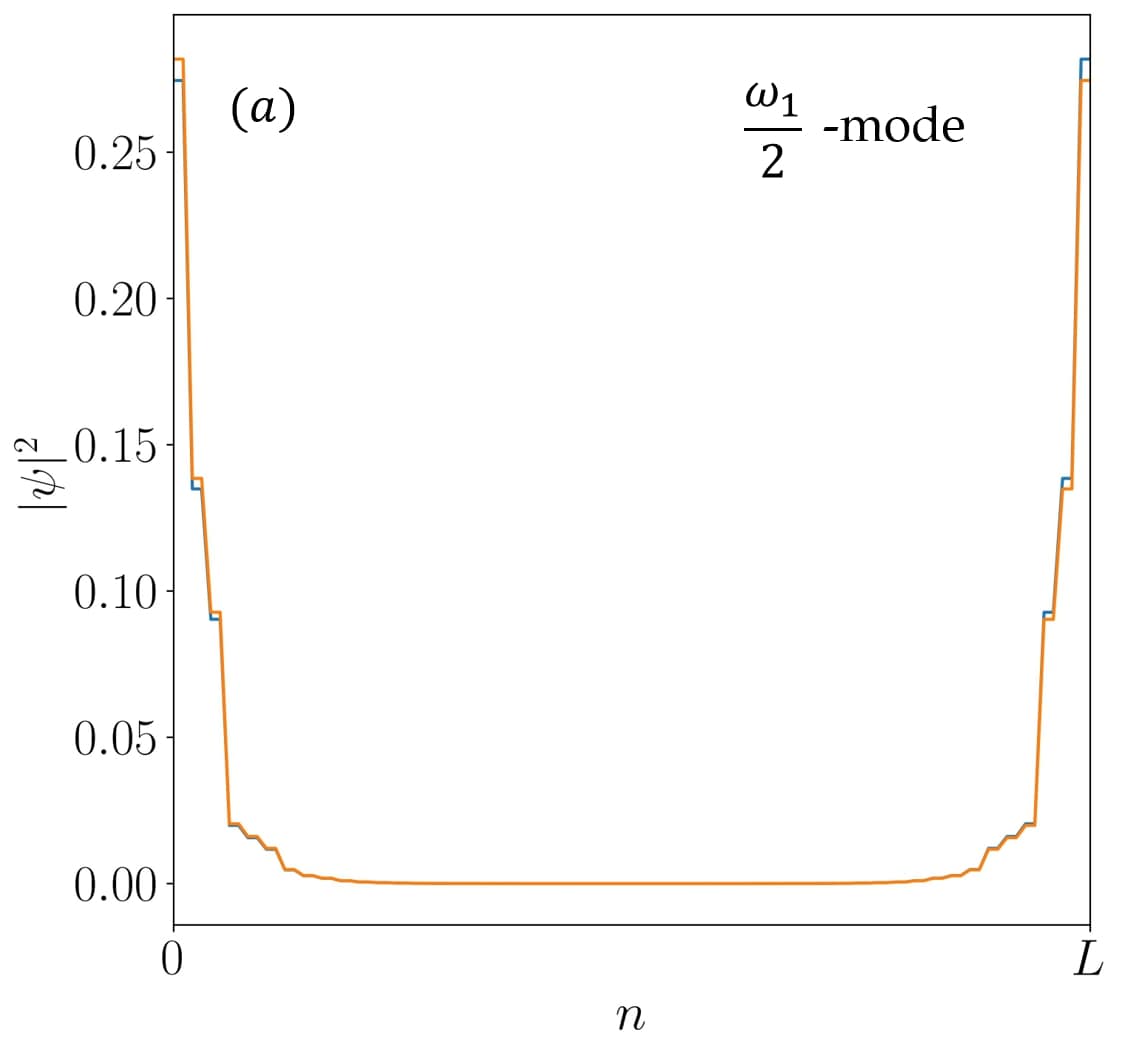}
\hfill
\hfill
\includegraphics[width=0.25\textwidth]{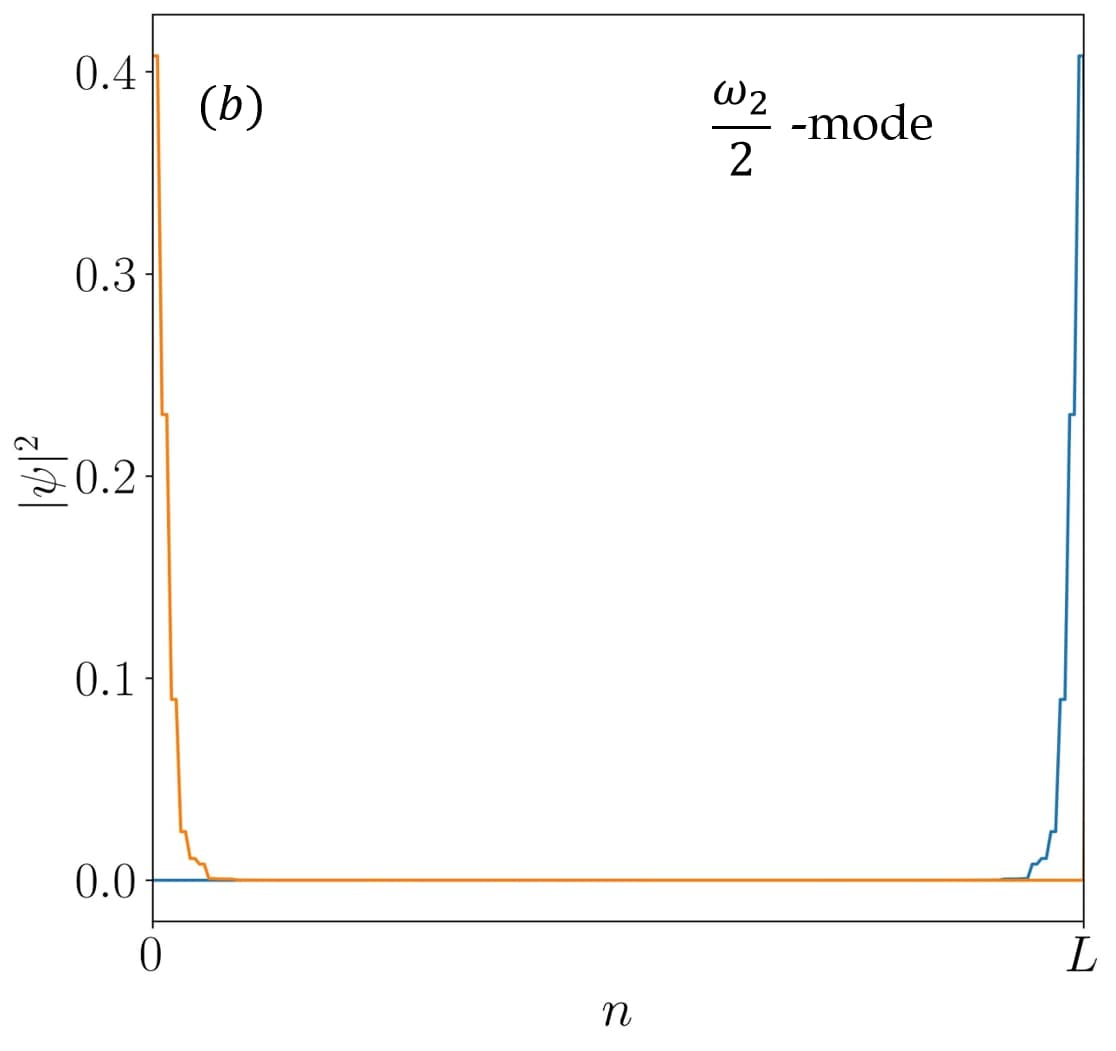}
\hfill}
\caption{Panel (a) and (b) represent the probability distributions of the two distinct Majorana states appearing at energy, $E=\omega_1/2$ and $E=\omega_2/2$ respectively corresponding to a fixed magnetic field, say, $B=1.5$. The ratio of the two frequencies has been chosen as, $\omega_2/\omega_1 = \beta= \sqrt{5}+1/2.$ The strength of the two drives are fixed at, $B_1=B_2=1$.}
\label{Fig11}
\end{figure}
\textit{Incommensurate case:} On the other hand, corresponding to an incommensurate driving scheme, the interplay between the two mutually irrational drives can lead to the emergence of various types of $\pi$ Majorana modes, as evidenced by the presence of two distinct level crossings at energies, $E = \omega_1 /2$ and $E = \omega_2 /2$ (Fig.~\ref{Fig10}, where the gap closes at two different $k$ values, shown by the dashed circles). In an open chain, as soon as the time-dependent perturbation is switched on, these could give rise to the emergence of distinct localized edge states (shown in Figs.~\ref{Fig11}). Furthermore, the appearance of these unique $\pi$ modes can be regulated independently due to the fact that by keeping one drive constant while setting the other to be zero, it is possible to switch between different bulk gaps. For instance, starting with $B_2 = 0$ and $B_1  \neq 0$, the gap opens solely at $E=\omega_1/2$, and when $B_2 \neq 0$ and $B_1=0$, the gap opens at $E=\omega_2/2$. The existence and stability of the Majoranas have been affirmed through the analysis of the probability distribution corresponding to the eigenstates (Figs.~\ref{Fig11}).
\begin{figure}[t]
          \begin{subfigure}[t]{0.8\columnwidth}
         %\centering
         \includegraphics[width=0.9\columnwidth]{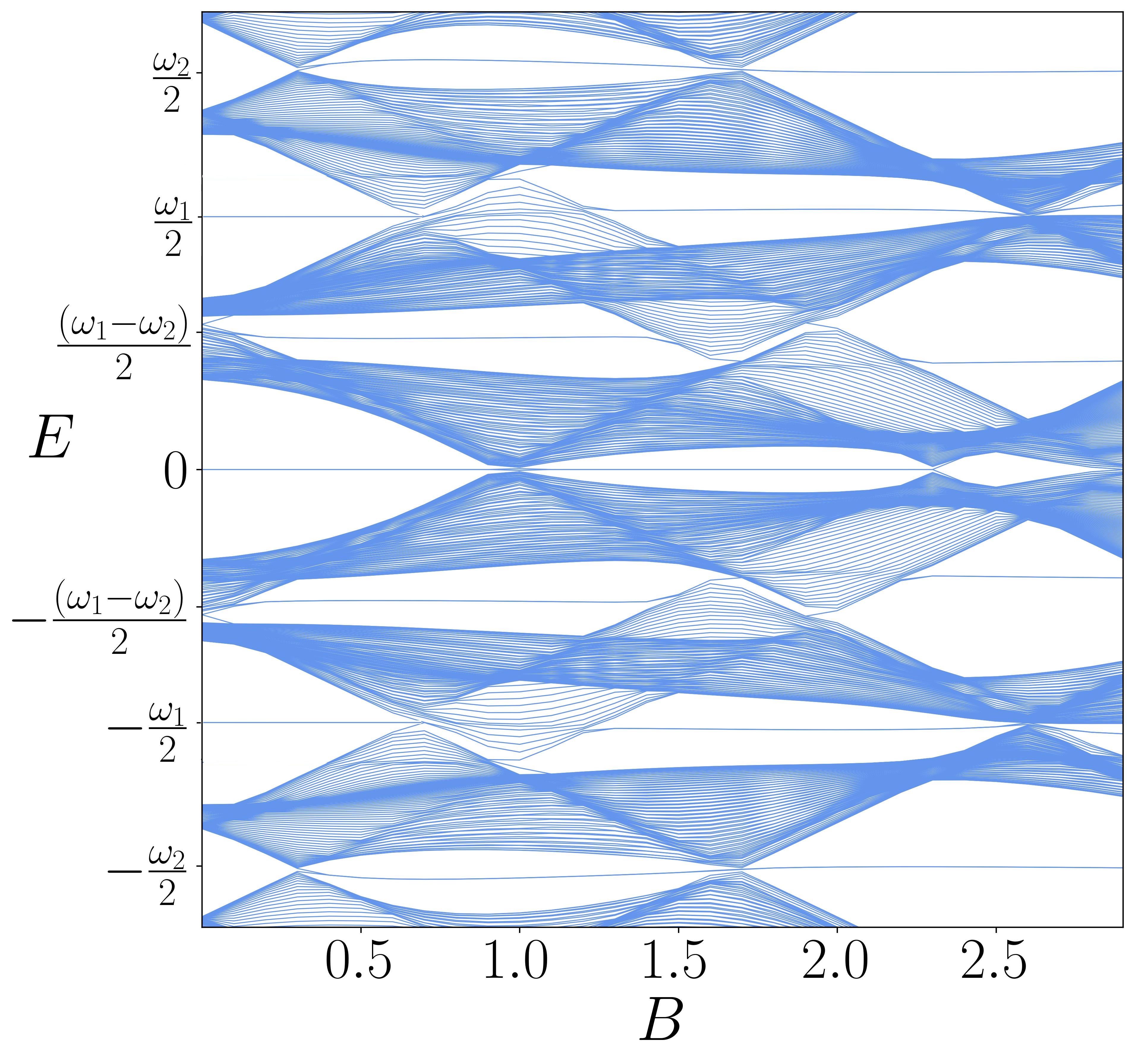}
         \captionlistentry{}
         \label{10.1}
     \end{subfigure}
\caption{{The figure represents the real-space quasi-energy spectrum in the many-mode Floquet formalism corresponding to an incommensurate multi-frequency driving protocol plotted as a function of $B$, for a fixed frequency, say, $\omega_1=3$ ($T=2.09$). The ratio of the two frequencies has been chosen as, $\omega_2/\omega_1 = \beta= (\sqrt{5}+1)/2.$ The strength of the two drives are fixed at, $B_1=1,B_2=1.5$.}} 
 \label{Fig12}
\end{figure}
\par Further in Fig.~\ref{Fig12}, we illustrate the real-space quasi-energy spectra corresponding to three replicas in each of the frequency directions plotted as a function of $B$ with fixed frequency, say $\omega_1=3$ ($T=2.09$). The different $\pi$ energy modes are clearly visible at non-zero energies along with the zero-energy modes. Additionally, we observed that when the strengths of the two drives are equal, there can be at most two localized edge modes along with the zero energy modes at energies $E=\omega_1/2$ and $E=\omega_2/2$ respectively. Whereas, with $B_1 \neq B_2$, another variant of $\pi$ energy mode appears at energy, $E = (\omega_1 - \omega_2)/2$. 
\par
\begin{figure}[t]
    \begin{subfigure}[b]{\columnwidth}
         \includegraphics[width=\columnwidth]{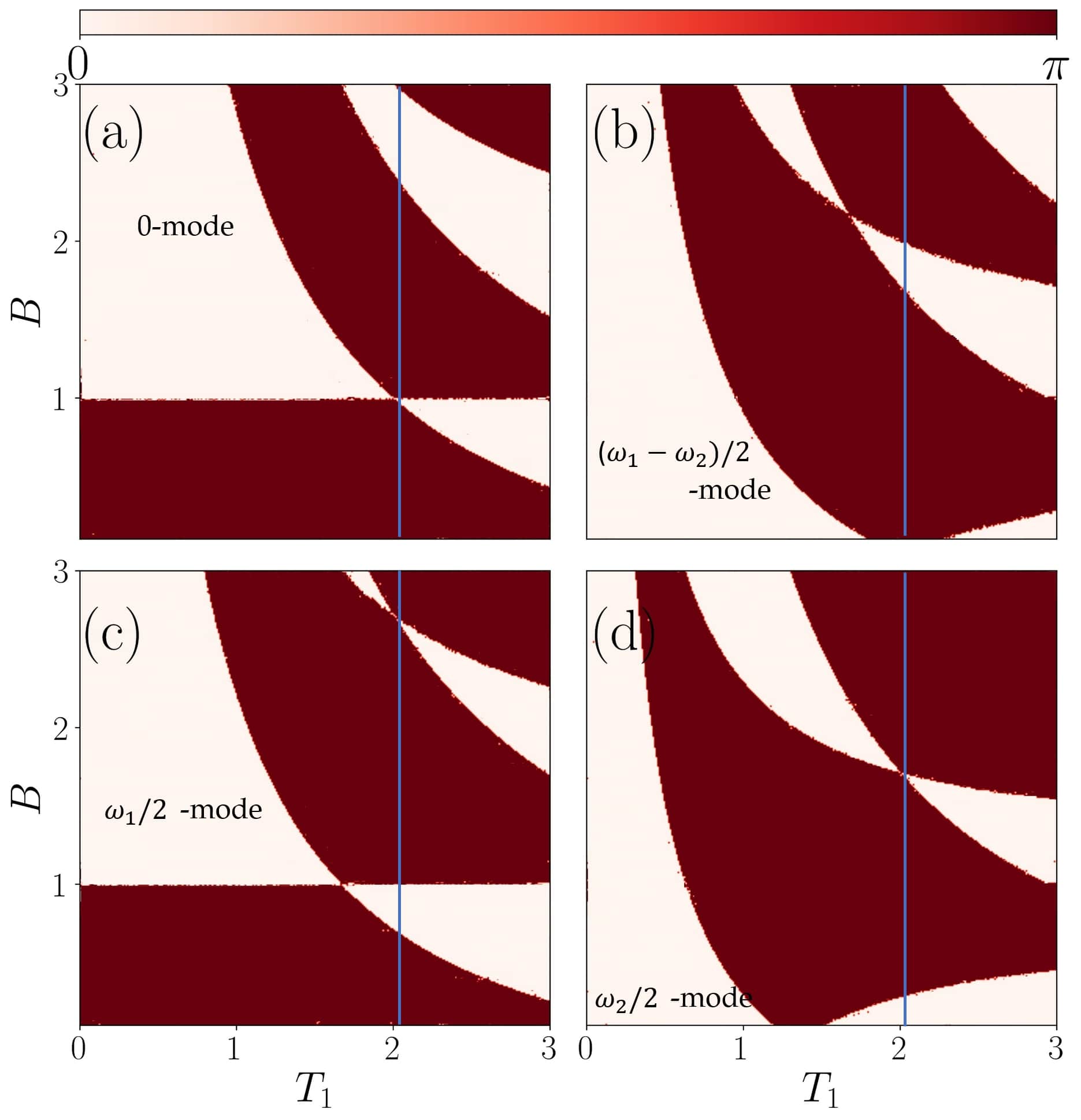}
        %  \caption{}
         \label{Fig1}
     \end{subfigure}
\caption{{The figure depicts topological phase diagrams plotted on the basis of the total Berry phase filled upto states $E=0$ (panel (a)), $E=(\omega_1 - \omega_2)/2$ (panel (b)), $E = \omega_1/2$ (panel (c)) and $E=\omega_2/2$ (panel (d)) respectively. the strength of the two drives are chosen as, $B_1 =1, B_2 = 1.5$.}} 
\label{Fig13}
\end{figure}

 However, unlike the situation in the commensurate case, we cannot establish a winding number because of the inability to define a stroboscopic time evolution operator. Instead, similar to the Shirley-Floquet approach, different Majoranas can be characterized using the Berry phase. In Fig.~\ref{Fig13}, we have plotted the topological phase diagram in the $B-T_1$ plane concerning the Berry phases corresponding to different Majorana modes by taking the cumulative sum of the phases corresponding to the bands below energies, $E=0, \omega_1 /2$, $\omega_2 /2$ and $(\omega_1 - \omega_2)/2$ respectively. Further, one can confirm the existence of bulk-edge correspondence by following the vertical lines ($T_1 = 2.09$) in the phase diagrams shown in Fig.~\ref{Fig13}. It is noteworthy that there are regions in the parameter space where all these Majoranas can coexist independently of each other. As a consequence, this could offer potential applications for tunable control over accessing localized edge states. For instance, 
Instead of relying on several static topological superconducting wires, one can dynamically produce multiple Majoranas at various locations of the same network. This presents an intriguing avenue for future exploration, potentially opening up opportunities to investigate diverse Floquet characteristics.
\par At this stage, It is crucial to investigate whether the many-mode Floquet theory discussed via Eq. \ref{MMFT_compact_matrix} that corresponds to the incommensurate case can be used for the commensurate driving scheme as well. To answer the question, the same formalism can not be carried over the commensurate case. Generally, solving the commensurate case using Eq. \ref{MMFT_compact_matrix} can provide results comparable to those in Fig. \ref{Fig9}, which is shown for the commensurate case using the traditional method, albeit only for the high frequency regime. Moreover, if there exists an additional phase difference between the two drives, the results from Eq. \ref{MMFT_compact_matrix} and the conventional method show notable dissimilarities, that would depend upon the phase difference \cite{validityofmultimode}. This can be understood by the fact that unlike the incommensurate scenario, where there is no global time periodicity, in the commensurate driving scheme, however, the system retains its global time periodicity. Thus, the periodicity essentially compresses the frequency space into a cylindrical structure, where certain frequency lattice sites, for example, $\vec{m}$ and $\vec{m} + \vec{l}$ (where $\vec{l} = (n_1,n_2)$) become equivalent. As a result, the quasienergies in the commensurate case reflect a compact cylinderical system with circumference $|\vec{l}|$, and hence distinct from the extended structure in the incommensurate case. Moreover, in the incommensurate case, a conventional choice of a finite basis set is typically employed for the numerical diagonalization of the Floquet Hamiltonian, as outlined in \cite{validityofmultimode}. For the commensurate driving scheme, however, the diagonalization involves specific choices of basis sets that correspond to points within the compact cylinder \cite{floquetincommensurate5,validityofmultimode}.
\par Therefore, while Eq. \ref{MMFT_compact_matrix} provides an elegant solution for the incommensurate case, it fails being entirely adequate for the commensurate scenario. Thus in order to perform computation for the latter it is more suitable to use the conventional approach, such as the Floquet evolution operator. Additionally, the SF method can be applied too, which however necessitates the problem to be mapped onto a 1D tight-binding model with an inclusion of a next-nearest neighbor hopping term ($H_{\pm 2}$) that is going to distinguish the presence of the additional drive, as
explained in Ref. \cite{solin}.
\par Further, our insights that a Rashba nanowire proximitized with an $s$-wave superconductor under the application of periodic drive simulates rich non-trivial characteristics allow us to suggest a route for the experimental validation of our results. Specifically, our proposal requires strong spin-orbit coupling in a conducting nanowire. Numerous experimental and theoretical studies explore various aspects of spin-orbit coupling in wires \cite{so_coupling_in_wires1,so_coupling_in_wires2,so_coupling_in_wires3}. However, a more promising candidate is an InAs or InSb wire in the wurtzite structure, which are known to exhibit Rashba-type spin-orbit coupling \cite{inas_rashba}. InAs nanowires in proximity to superconductors, such as Nb and Al have been experimentally studied and are known to form highly transparent interfaces for electrons enabling the induction of a large superconducting gap \cite{inas_rashba_proximity}. Additionally, an in-plane magnetic field can create a gap in the spectrum at zero momentum, effectively eliminating the spin degeneracy. Due to the significant difference in the $g$-factors between Nb $(g \sim 1)$ and InAs $(g_{\text{InAs}} \sim 35$ and $g_{\text{InSb}} \sim 50 )$ \cite{g_factor}, the in-plane magnetic field can open a large Zeeman gap without significantly suppressing the SC gap in Nb. Additionally, the spin-orbit coupling strength can be related to the experimentally measured length scale $\lambda_{SO} \sim 200$ nm corresponding to a Rashba parameter, $\alpha = 0.2 \, \text{eV} \cdot \text{\AA}$ \cite{expt_length_scale_rashba}. This implies a spin-orbit energy scale $E_{SO} \approx \frac{\alpha^2 m^{*}}{2 \hbar^2} \approx 50 \, \mu \text{eV}$, where $m^{*} = 0.015 \, m_e$ is the effective mass of electrons in InAs. Consequently, the other parameters can be assumed to have values such that $E_{SO} \simeq t \simeq 50 \, \mu \text{eV}$. Therefore, for a magnetic field of 0.15 T, the non-trivial characteristics can emerge at frequencies beyond approximately $\omega \simeq 10-20$ GHz, as indicated by the expression of critical frequencies outlined in Sec.~\ref{sec:level2}.
\par Now, in order to incorporate periodic driving, one can consider a waveform in the form of a bi-harmonic excitation as shown in Ref.~\cite{two_tone_drive_expt}. Further, the rationality between the two drives can be altered by tuning the phases between the two harmonics.
\par Finally, To provide evidence of a topological phase transition, one can detect spin or charge reversal at the phase boundaries using spin-polarized scanning tunneling microscopy (STM) \cite{stm_rashba1,stm_rashba2}. In this technique, a current is injected into the lowest energy bands of the material. The current flow is influenced by the polarization of the STM probe: in the trivial phase, current flows only if the polarization of the probe aligns with the spin orientation of the material; otherwise, it does not flow.  Conversely, in the topological phase, the situation is reversed, and the current flow behavior changes depending on the polarization characteristics of the probe.

\section{\label{sec:level5}Conclusion}
In summary, this paper explores a feasible topological superconductor model involving a Rashba nanowire subjected to multiple driving forces, aiming to generate numerous and novel Majorana modes. By leveraging the topological phase boundaries of the undriven model, we confirm the presence of distinct non-trivial characteristics at certain frequency limits. For instance, corresponding to the static trivial (topological) limits, the Majorana zero modes appear only at low (high) frequency regimes. Additionally, the non-trivial nature of the system is validated through the computation of several bulk invariants, such as the Berry phase associated with the Shirley-Floquet formalism. Further, the periodic drive induces longer-range interactions, leading to the generation of multiple Majorana modes where there is a possible alteration of the symmetry classification on basis of certain symmetry restoration. By exploiting the symmetries associated with the Floquet stroboscopic operator, we have been able to classify the effective Hamiltonian via a pair of bulk invariants that uniquely predict the emergence of both zero and $\pi$ Majorana modes. Later, the inclusion of an additional drive leads to intriguing results that could have potential applications in quantum computation. Within a commensurate driving scheme, the ability to dynamically manipulate the generation and stability of the edge modes are feasible by carefully choosing a suitable multi-frequency driving protocol. This results in a diverse range of edge states that exhibit varying degrees of localization, spanning from highly localized to fully delocalized ones. Moreover, more intricate driving protocols tend to populate the spectrum more densly and diminish the size of the quasienergy gaps, leading to the delocalization of the edge modes. Thus, by switching between various driving protocols, one can gain dynamical control over the presence and robustness of Majorana modes. However, in an incommensurate driving scheme, the absence of global time periodicity prevents the definition of usual stroboscopic evolution operator, limiting our analysis to the frequency space. Nevertheless, similar to Shirley's formalism, we have effectively addressed the problem in an extended frequency domain. This leads to the emergence of distinct Majorana modes not located at $E=0$, rather at $E= \omega_1 /2,$ and $ E=\omega_2 /2$. As a result, the simultaneous existence of time-quasiperiodic Majoranas could offer opportunities for exercising control over quantum computation by accessing multiple Majoranas at various locations within the same network.

\end{document}